\documentclass{article}

\usepackage{arxiv}

\usepackage{latexsym}
\usepackage{graphicx}
\usepackage{multicol,multirow}
\usepackage{amsthm, amsmath,bbm,amssymb,amsfonts}
\usepackage{mathrsfs}
\usepackage{algorithmic}
\usepackage{algorithm}
\usepackage{rotating}
\usepackage{appendix}
\usepackage[authoryear]{natbib}
\usepackage{ifpdf}
\usepackage[T1]{fontenc}
\usepackage{times}
\usepackage{sourcesanspro}
\usepackage{newtxmath}
\usepackage{textcomp}%
\usepackage{xcolor}%
\usepackage{hyperref}
\usepackage{caption}
\usepackage{subcaption}
\usepackage{mathtools}
\usepackage{tabularx}
\usepackage{tikz}
\usetikzlibrary{shapes.geometric, arrows}
\usepackage{placeins}

\newcommand\iidsim{\stackrel{\mathclap{iid}}{\sim}}
\usepackage[utf8]{inputenc} 
\usepackage{url}            
\usepackage{booktabs}       
\usepackage{nicefrac}       
\usepackage{microtype}      
\usepackage{lipsum}		
\usepackage[shortlabels]{enumitem}
\usepackage{dsfont}

\newcommand{\bs}{{\bf s}}

\usepackage{makecell}

\usepackage{natbib}

\tikzstyle{block} = [rectangle, 
minimum width=6cm, 
minimum height=1cm,
text width=5cm,
draw=black]
\tikzstyle{title} = [rectangle, 
minimum width=6cm, 
minimum height=2cm,
draw=black]
\tikzstyle{arrow} = [thick,->,>=stealth]

\theoremstyle{plain}

\theoremstyle{remark}

\title{Tracing the impacts of Mount Pinatubo eruption on global climate using spatially-varying changepoint detection}

\author{Samantha Shi-Jun\
	University of Illinois, Urbana-Champaign\\
	Champaign, Illinois\\
	Sandia National Laboratories\\
	Albuquerque, NM \\  \\
	\And
	\href{https://orcid.org/0000-0002-5239-5185}{ \includegraphics[scale=0.06]{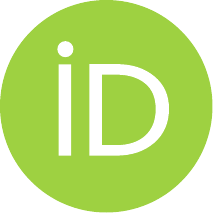}\hspace{1mm}Lyndsay Shand} 
	\thanks{corresponding author, lshand@sandia.gov} \\
	Sandia National Laboratories\\
	Albuquerque, NM \\	\\
	University of Illinois, Urbana-Champaign\\
	Champaign, Illinois\\
	\And
	Bo Li\\
	University of Illinois, Urbana-Champaign\\
	Champaign, Illinois\\
	Sandia National Laboratories\\
	Albuquerque, NM \\
}

\renewcommand{\shorttitle}{\textit{arXiv} Template}

\begin{document}
\maketitle

\begin{abstract}
Significant events such as volcanic eruptions can have global and long-lasting impacts on climate. These global impacts, however, are not uniform across space and time. Understanding how the Mt. Pinatubo eruption affects global and regional climate is of great interest for predicting the impact on climate due to similar events as well as understanding the possible effect of the Stratospheric Aerosol Injections proposed to combat climate change. While many studies illustrated the impact of the Pinatubo eruption on a global scale, studies at a fine regional scale are scarce. 
We propose a Bayesian spatially-varying changepoint detection and estimation method to trace the impact of Mt Pinatubo eruption on regional climate. 
Our approach takes into account the diffusing nature and spatial correlation of the climate changes attributed to the volcanic eruption.
We illustrate our method and demonstrate its advantages over an existing changepoint detection method through simulations. Finally, we apply our method to monthly stratospheric aerosol optical depth and surface temperature data from 1985 to 1995 to detect and estimate changepoints following the 1991 Mt. Pinatubo eruption. Our results quantitatively characterize the spatial pattern of the eruption's impact on regional climate, complementing the previous studies on the global impact of the Pinatubo eruption. 
\end{abstract}


\section{Introduction}\label{sec:intro}
The Mount Pinatubo eruption in June 1991 is the largest volcanic eruption in recent history that injected nearly 20 megatons of sulfur dioxide into the atmosphere. The aerosol cloud resulting from the eruption encircled the globe for a few weeks, and global changes to the atmosphere, including a decrease in surface temperature in the northern hemisphere of about 0.5$^\circ$ C, were observed up to two years after the event \citep{self1993}. 
Studying the impact of volcanic eruptions is of high interest to the climate community for two main reasons. On the one hand, it helps improve predictions of the anticipated impacts from new volcanic activities or other localized events such as wildfires. On the other hand,  volcanic eruptions serve as natural examples of Stratospheric Aerosol Injection (SAI), a controversial proposed solar climate intervention to help cool the earth and reduce the impact of global warming through a global dimming effect \citep{robock}. Thus, a more comprehensive understanding of the impact of the Pinatubo eruption can shed new light on this radical technology against climate change.

Scientists have extensively studied how atmospheric properties were altered by the Mt. Pinatubo eruption. They found that the radiative forcing changes lasted for about three years \citep{stenchikov2009}, leading to observed heating in the stratosphere of 2-3K \citep{lm1992} and cooling of global surface temperatures of about 0.4 K \citep{dc1992, thompson2009}. Although there is a consensus that the eruption ``caused" these global atmospheric changes, most published studies provided only qualitative evidence on a global scale. 
In this study, we attempt to quantitatively trace the impact of the Pinatubo eruption on aerosol optical depth and temperatures at a finer scale.

Since climate impacts from a volcanic eruption or a general SAI event are expected to cause sudden changes in regional climate, it is natural to use changepoints to trace the impact of an eruption at a regional scale. Changepoint detection methods aim to identify significant shifts in the underlying distribution of the data and have been employed to understand the impact of significant events on the environment. For instance, \cite{hallema2017assessment} used changepoint analysis to assess the impacts of wildland fires on watershed annual water yield,  \cite{robbins2011changepoints} developed a new changepoint detection method to identify changes in the tropical cyclone record in the North Atlantic Basin along with climate change over the period 1851–2008, and \cite{tucker2023elastic} found evidence of  increased stratospheric temperature after the Mt. Pinatubo eruption using an epidemic changepoint model. 

There have been many developments in changepoint detection in both univariate and multivariate settings, including two review articles \citep{reeves2007review, aminikhanghahi2017survey} and references therein. In particular, methods such as pruned exact linear time (PELT) algorithm \citep{killick2012optimal} and product partition model (PPM) \citep{barry1992product} have gained considerable popularity in the changepoint literature. In situations where time series are indexed by spatial locations, the change behavior of time series that are geographically nearby is likely to be similar due to spatial correlation. For such data, the direct application of traditional changepoint detection methods to each time series might overlook this spatial feature and is inadequate to exploit spatial correlation to improve changepoint detection and estimation. 

To address this limitation, \cite{majumdar2005spatio} employed a Bayesian approach to detect various types of changepoints by incorporating spatial correlation. Additionally, \cite{xuan2007modeling} extended the PPM to account for the dependency structure in multivariate time series using sparse Gaussian graphical models. While these methods can handle spatial dependence among time series, they assume a common changepoint across all spatial locations, imposing a strong limitation on their application.
Recently, a spatially-varying changepoint model was proposed by \cite{berchuck2019spatially} to monitor glaucoma progression in visual field data, where multivariate conditional autoregressive  priors are used to model the spatial dependencies in the changepoints and the mean parameters. Their model assumed the data is temporally independent. Another related work is \cite{wang2023asynchronous}, who developed a spatially-varying changepoint estimation method for functional time series observed over a spatial domain.

The changes in regional climate caused by the Pinatubo eruption or a general SAI event are expected to exhibit similarity at nearby locations due to spatial dependency. However, these changes are anticipated to occur at different times due to the diffusion of the injected aerosols across the globe and the atmospheric circulation. For particular variables such as the global aerosol optical depth (AOD) that measures the extinction of the solar beam by dust and haze, we expect that the time of the change will increase as the spatial distance from the event location increases, again due to the diffusion of aerosols. 
These unique characteristics of our problem render existing methods inadequate for our application. We, therefore, propose a novel space-time changepoint detection method that can identify when, where, and to what extent climate impacts have occurred following the 1991 eruption of Mt. Pinatubo while respecting the particular features of the SAI event. 

Our approach departs from conventional methods by modeling changepoints as a spatial process and further allowing time-after-event of the changepoints to increase with the spatial distance from the event origin. By taking into account the spatial correlation of the data and the anticipated diffusion of the observed impact from the event location, our method is demonstrated to be more effective in capturing the spatial patterns of climate impacts associated with the SAI events. Since our focus is solely on tracing the impact of the Mt. Pinatubo eruption, we consider at most one change after the eruption. 
Compared to \cite{majumdar2005spatio}, our method extends their model by allowing for spatially-varying changepoints. Furthermore, our approach performs both detection and estimation of changepoint at each location rather than only the estimation. Lastly, while most traditional methods focus on the mean shift, our method can detect changes in either the mean or variance, thus offering greater flexibility.

The rest of this paper is organized as follows. Section \ref{sec:data} describes the data and presents exploratory data analysis results. Section \ref{sec:method} introduces the model formulation and sampling procedure. In Section \ref{sec:simstudy}, we conduct extensive simulation studies to evaluate the performance of our proposed method. Section \ref{sec:results} demonstrates our methodology on AOD and surface temperature data observed around the time of the Mt. Pinatubo eruption. Finally, in Section \ref{sec:disc} we provide a brief summary and discuss potential future work.

\section{Data}\label{sec:data}

As mentioned in Section \ref{sec:intro}, notable changes in aerosol optical depth (AOD) and surface temperatures caused by the Pinatubo eruption have been documented. While aerosols rapidly spread around the globe and experienced a near immediate and steep increase in magnitude following the eruption, changes in surface temperature were more subtle and took years to manifest.  
We consider the stratospheric AOD that measures the extinction optical thickness at 550nm\footnote{referred to as``TOTEXTTAU'' in \cite{totexttau}}, and the surface temperature\footnote{referred to as ``tavg1\_2d\_slv\_Nx'' in \cite{surface_temp}}, to evaluate the impact of the eruption. 
We obtained both data sets from the Modern-Era Retrospective analysis for Research and Applications, Version 2 (MERRA-2)\footnote{\url{https://gmao.gsfc.nasa.gov/reanalysis/MERRA-2/}}. 

\begin{figure}[!ht]
\centering
  \includegraphics[width=14.2cm]{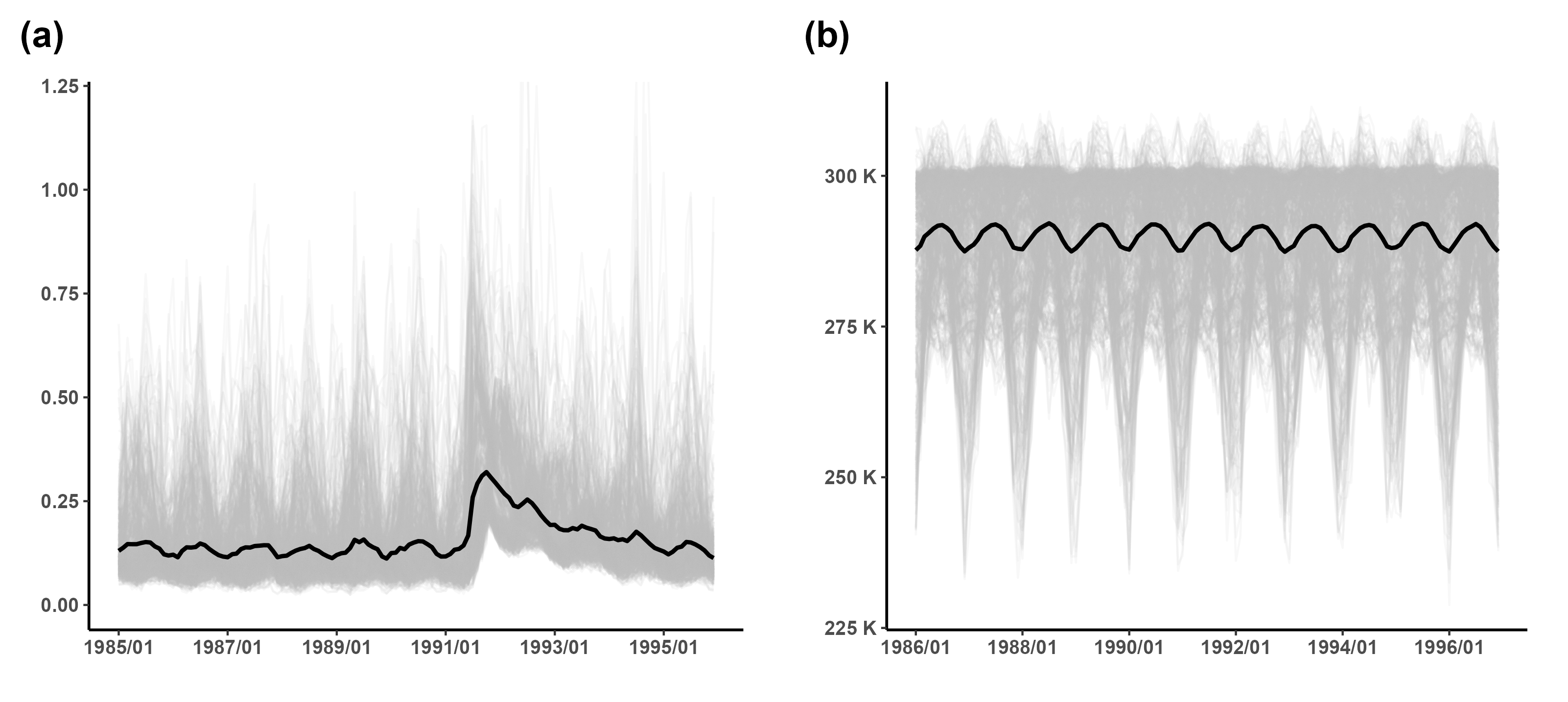}%
  \caption{\label{fig:raw_data} Time Series of (a) stratospheric AOD data and (b) surface temperature data before preprocessing. Each grey curve represents a time series at a grid point. The black curve represents the global average.}
\end{figure}

The monthly AOD data ranges from January 1985 to December 1995, covering 132 time points in total. The data is over a $48\times24$ grid and covers the entire globe. Following \cite{self1993}, we exclude latitudes above $60^\circ$N and below $60^\circ$S in our analysis, resulting in $48\times 16 = 768$ spatial locations. Figure \ref{fig:raw_data}(a) shows the AOD data on these 768 spatial grids. In general, the aerosol data exhibits a clear jump after the eruption, followed by an approximately linear decay in time toward its previous level. However, not all locations precisely agree on the jump time, likely due to the aerosol diffusion process. To enhance our analysis, we first remove the seasonality by applying Seasonal-Trend decomposition using LOESS (STL) \citep{rb1990stl} to AOD data at each location. Then, we estimate the temporal trend as a linear function of time using the historical data up to June 1991, the month of the eruption. If the estimated trend parameter is statistically significant at a 0.05 level, we remove the estimated trend for both pre- and post-eruption data. Assuming the trend between the month of eruption and the actual changepoint remains the same as in the pre-eruption period, this procedure roughly centers the time series before the changepoint so a constant mean can be assumed. Finally, we normalize the time series at each location by calculating the sample variance using data before June 1991 and use this number to achieve approximately constant variance across all locations. 

The monthly surface temperature data, shown in Figure \ref{fig:raw_data}(b), also spans from January 1985 to December 1995 and is on the same $48\times 24$ grid as the AOD data. In contrast to the AOD data, the impact of volcanic eruption on temperatures is indiscernible through simple visualization, suggesting a more subtle influence on temperatures, if any. When examining how temperature responds to a volcanic eruption, it is customary to study the temperature zonal mean indexed by latitude \citep{robock1995volcanic, stenchikov2002arctic, gao2008volcanic}. Hence, we average temperatures over longitude for each of the 24 latitude bands given in the data and conduct the analysis on the latitudinal means. This reduces the number of observations to $24$, with each latitude representing a latitude band covering $7.5^\circ$. We employed the same procedure as for the AOD data to detrend and normalize temperature data. The processed zonal mean temperatures can be seen in Figure \ref{fig:surface_lat_mean}. 

\begin{figure}[!ht]
\centering
  \includegraphics[width=14.2cm]{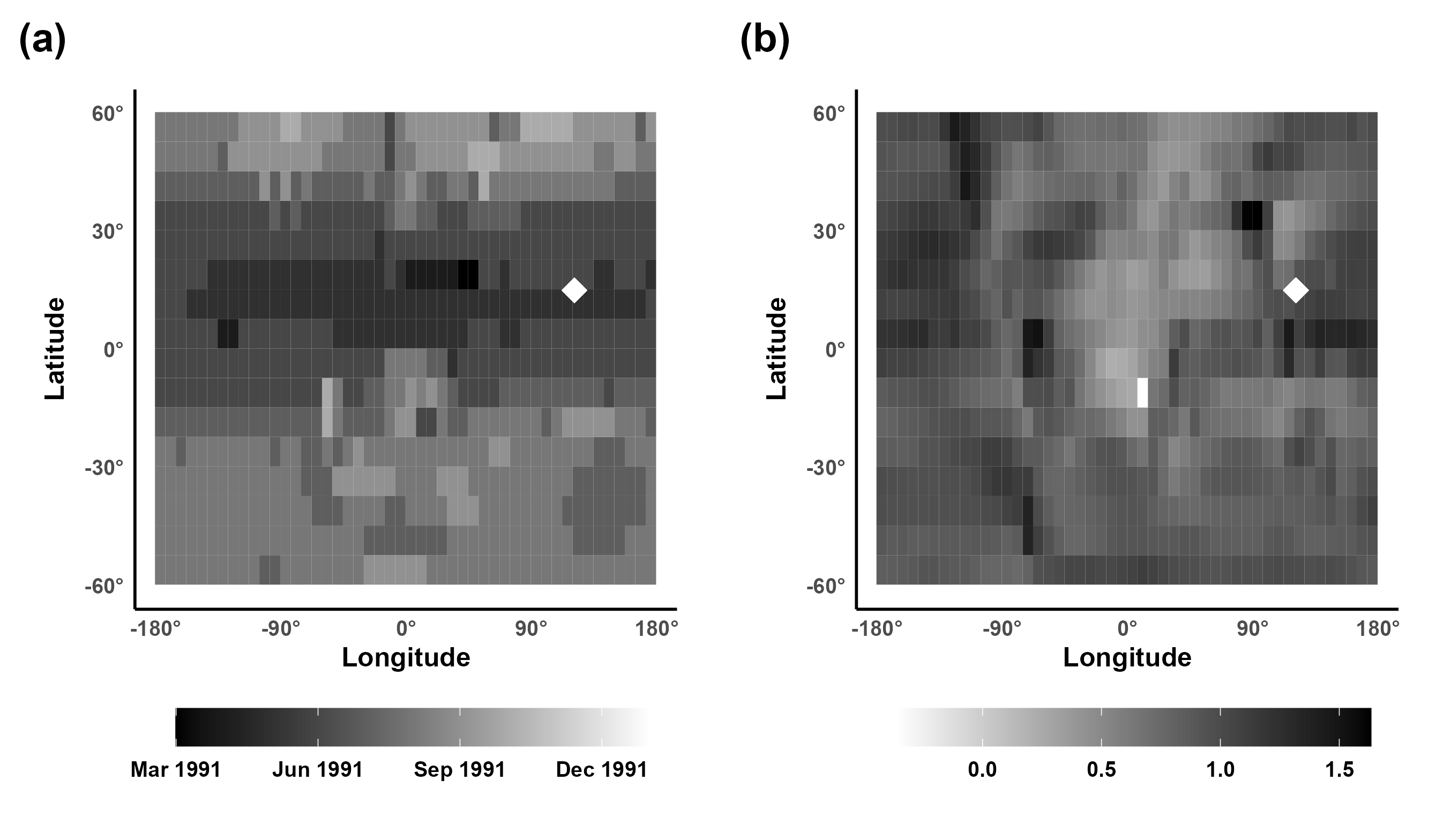}%
  \caption{\label{fig:heatmap_eda} (a) Heatmap of detected changepoints for AOD. Darker color indicates earlier change. (b) Heatmap of estimated change amount. Darker color indicates stronger magnitude of mean shift. The white diamond marks the location of Mt. Pinatubo. }
\end{figure}

To evaluate our conjecture that changes in these climate variables vary spatially, we conducted exploratory data analysis on the global AOD data. At each grid location, we applied the Bayesian information criteria-based changepoint detection procedure, which is a well-established method for climate series \citep{reeves2007review}.
Figure \ref{fig:heatmap_eda} illustrates the heatmap of detected changepoints and estimated mean shifts at each grid location. Overall, the changepoints appear to be spatially-varying and clustered, as does the magnitude of change. Although a few locations show changepoints prior to 1991/06 that are apparently not due to the Mt. Pinatubo eruption, in general, the changepoints appear to spread latitudinally from the source of the eruption. This spatial pattern is consistent with what we expect from an eruption event, given the influence of strong zonal winds in the lower stratosphere that circulate the globe \citep{robock1983circumglobal}. These results not only verify our conjecture but also provide guidance for the development of our methodology.

\section{Method}\label{sec:method}
We propose a Bayesian hierarchical model to detect and estimate spatially-varying changepoints for spatiotemporal data. We model the likelihood of the data in the first level, provide the prior for the latent processes in the second level, and close the hierarchy by specifying priors and hyperpriors for all unknown parameters in the third level. We then discuss the challenges and the schemes of the Markov Chain Monte Carlo (MCMC) sampling algorithm used to estimate the proposed model.  

\subsection{Changepoint detection and estimation}

Let $Y(\mathbf{s},t)$ denote the observed climate variable at location $\mathbf{s} \in \mathbb{S}^2$ and time $t\in \mathbb{N}$, $N$ be the total number of spatial locations, and $M$ the total time points. We propose the following model for the spatiotemporal process $Y(\mathbf{s},t)$ with possible mean and variance changes in time:

\noindent\textbf{Level 1 } The likelihood of the $Y(\mathbf{s},t)$ process: 
\begin{align} \label{eq:1}
Y(\mathbf{s},t) &= \begin{cases} \mu_1(\mathbf{s},t) + U(\mathbf{s},t) + \epsilon_1(\mathbf{s},t), &t \leq \tau(\mathbf{s}) \\ \mu_2(\mathbf{s},t) + U(\mathbf{s},t) + \epsilon_2(\mathbf{s},t), &t > \tau(\mathbf{s}), \end{cases}
\end{align}
where $\tau(\mathbf{s})$ is the changepoint at location $\bs$, $\mu_1(\mathbf{s},t)$ and $\mu_2(\mathbf{s},t)$ are the mean functions before and after changepoint,  $U(\mathbf{s},t)$ is a zero mean space-time correlated error process, and   $\epsilon_1 \iidsim N(0,\sigma_1^2)$ and $ \epsilon_2 \iidsim N(0,\sigma_2^2)$ are the before and after changepoint measurement errors. When $\tau(\mathbf{s}) = M$, it indicates the absence of changepoint at location $\bs$ as the entire process is governed by a single model. To keep our model parsimonious and avoid potential identifiability issues, we assume a common spatiotemporal process $U(s,t)$ both before and after changepoint. This ensures that all changes will be attributed to mean and/or variance of measurement errors. This assumption also allows us to take advantage of the possible space-time separable covariance structure to facilitate the computation for spatiotemporal data. While \cite{majumdar2005spatio} allows for a change in the covariance structure by modeling the pre-changepoint error process as $U(\bs,t)+V(\bs,t)$ and  the post-changepoint error process as $U(\bs,t)+W(\bs,t)$, 
their model assumes $\tau(\bs)$ as well as $\mu(\bs,t)$ to be constant as a sacrifice to ensure identifiability. Since the primary interest for our data is to identify the spatially-varying changepoints while respecting the spatiotemporally-varying mean, we opt for a model that offers this flexibility.

As mentioned in Section \ref{sec:data}, the data after preprocessing can be assumed to have a constant mean prior to the changepoint. Based on Figure \ref{fig:raw_data}(a), we assume a linear trend in time after changepoint. We also observe from Figure \ref{fig:heatmap_eda} that the change amount not only varies spatially but also tends to be clustered in space. To account for these features, we model the pre and post-changepoint mean processes, $\mu_1$ and $\mu_2$ respectively, as 
\begin{align}
    \mu_1(\mathbf{s},t) &= \alpha_0 \label{eq:RE1}\\
    \mu_2(\mathbf{s},t) &= \alpha_0 + \gamma_0(\mathbf{s}) + \gamma_1(\mathbf{s})\cdot(t-\tau(\mathbf{s})) \label{eq:RE2}\\
    &= \alpha_0 + (\gamma_{0F} +  \gamma_{0R}(\mathbf{s})) + (\gamma_{1F} +\gamma_{1R}(\mathbf{s}) )\cdot(t-\tau(\mathbf{s})) \label{eq:RE3}.
\end{align}
In (\ref{eq:RE1}) and (\ref{eq:RE2}), $\alpha_0$ is the constant parameter representing the average global mean before the changepoint, $\gamma_0(\bs)$ is the mean shift at the time of the changepoint at location $\bs$, and $\gamma_1(\bs)$ represents the slope after changepoint at locations $\bs$. We further decompose $\gamma_0(\bs)$ and $\gamma_1(\bs)$ into a fixed and random components in  (\ref{eq:RE3}). The fixed components $\gamma_{0F}$ and $\gamma_{1F}$ can be interpreted as the average mean shift and the average post-changepoint temporal trend across all locations that have a changepoint. The parameters $\gamma_{0R}(\bs)$ and ${\gamma_{1R}(\bs)}$ serve as spatial random effects to allow the mean shift and trend to vary spatially. To respect the spatial correlation of the magnitude of change and of the post-changepoint trend, we model $\gamma_{0R}(\bs)$ and ${\gamma_{1R}(\bs)}$ as a spatially correlated process. 

Let $\bs_0$ denote the location at which the change is first observed among all locations where we attempt to detect the changepoint, i.e.,  $\bs_0 = \arg \min_\bs \tau(\bs)$. If the influence of the volcanic eruption on the variable of interest diffuses from the event origin, $\bs_0$ is usually either the event origin or some location nearby depending on what locations we consider for changepoint detection.   
We model $\tau(\mathbf{s})$ as the sum of two terms: the changepoint at the location $\bs_0$, denoted by $\tau_0$, and the temporal lag between changepoints at $\mathbf{s_0}$ and $\bs$, denoted by 
$\Delta(\mathbf{s})$. 
Specifically, we have 
\begin{equation} \label{eq:cp}
\tau(\mathbf{s}) = \min\{M,\lfloor \tau_0 + \Delta(\mathbf{s}) \rfloor \}.
\end{equation}
In theory, the ``true'' changepoint process can be defined as $\tau_0 + \Delta({\bf s})$, where the temporal lags $\Delta({\bf s})$ are continuous and take values in $(0,\infty)$. However, the observed changepoint is discrete and we are mainly interested in changepoints between 1 to M, so we take the floor sign and cap the value of changepoints at $M$. We require $\Delta(\mathbf{s})\ge 0$ such that $\tau_0$ is the earliest changepoint. This ensures that all other changepoints detected are either at or after $\tau_0$ and thus not absolutely due to other unrelated events. We fix $\Delta(\mathbf{s}_0) = 0$ in our estimation because $\tau(\mathbf{s}_0) = \tau_0 = \min_{\bs}\tau(\bs)$,  as defined. 
Note that $\tau_0$ takes discrete values in $\{1, \ldots, M\}$ and $\tau_0 = M$ implies that there is no changepoint at any of the locations. There are two main advantages of separating $\tau_0$ from $\tau(\mathbf{s})$. First, it ensures that the changepoint at any location $\mathbf{s} \neq \mathbf{s}_0$ occurs strictly after the changepoint at $\mathbf{s}_0$, as any changepoint prior to $\tau_0$ would be due to another event that is not of our interest. 
Second, we can model $\Delta(\mathbf{s})$ as a continuous process with a spatial correlation structure to reflect the spatial pattern observed in Figure \ref{fig:heatmap_eda}(a) while keeping $\tau_0$ as a discrete variable. 
The discrete parameter $\tau_0$ can help us easily test if it is the case that no changepoint ever occurs at any of the locations and incorporate prior knowledge about the possible range of date at which the change caused by the eruption was first observed.

We further assume $\boldsymbol{\Delta}= (\Delta(\bs_1), \ldots, \Delta(\bs_N))^T$ follow a log normal process with mean $\mathbf{X}\boldsymbol{\beta}$ and covariance matrix ${\bf\Sigma}_\Delta$,
where $\mathbf{X}$ is the $N\times 2$ matrix of spatial lags from each location $\mathbf{s}$ to the location $\bs_0$, with the $i$th row of $\boldsymbol{X}$ given by $( \hbox{lon}(\mathbf{s}_i)-\hbox{lon}(\mathbf{s}_0), \hbox{lat}(\mathbf{s}_i) -\hbox{lat}(\mathbf{s}_0))$, and $\boldsymbol{\beta}$ is the 2-dimensional vector with each component representing the rate of diffusion in longitude and latitude, respectively. 
Allowing $\Delta(\mathbf{s})$ to increase with the distance from $\bs_0$ in an anisotropic manner reflects the nature of the physical process shown in Figure \ref{fig:heatmap_eda} that the eruption diffuses at a different rate along the longitude and latitude, but in any direction, it takes a longer time for the eruption impact to reach a specific location that is further away from the event origin.

Let $\boldsymbol{\gamma}_{0R}$ and $\boldsymbol{\gamma}_{1R}$ be the vector of $\gamma_{0R}(\bs_i)$ and ${\gamma_{1R}(\bs_i)}$ for $i=1,\ldots, N$, respectively. Let $\mathbf{U}$ be the vector stacking up all $U(\bs_i, t_j)$ for $i=1,\ldots, N$ and $j=1,\ldots, M$, and let $\log(\boldsymbol{\Delta})$ be the vector of element-wise logarithms of $\boldsymbol{\Delta}$. In our next level of the hierarchical model, we provide priors for the latent processes in model (\ref{eq:1}).\\
\textbf{Level 2} Latent processes:
\begin{align*}
      \boldsymbol{\gamma}_{0R} &\sim N(\mathbf{0}, \mathbf{\Sigma}_{\gamma_0}),\\
    \boldsymbol{\gamma}_{1R} &\sim N(\mathbf{0},\mathbf{\Sigma}_{\gamma_1})\\
        \mathbf{U} &\sim N(\mathbf{0},\mathbf{\Sigma_U}),\\
        \log (\boldsymbol{\Delta}) &\sim N(\mathbf{X}\boldsymbol{\beta},\mathbf{\Sigma}_\Delta).
\end{align*}
For computational simplicity, we assume that $\boldsymbol{\Sigma_U}$ has a separable space-time covariance structure with an exponential covariance function in both space and time, i.e. $\boldsymbol{\Sigma_U} = \sigma^2_U\mathbf{R}(\phi_U) \otimes \mathbf{R}(\psi_U),$ where the $(i,j)$th element of $\mathbf{R}(\theta)$ is 
$R(\theta)_{ij}= \exp(-\theta h)$ for a distance $h$ either in space or time,  
and $\otimes$ is the Kronecker product. Since our data spans the entire globe, we employed the great circle distance as the distance metric for the spatial correlation matrix. While a positive definite covariance function defined on Euclidean space may not necessarily be valid on spheres, the exponential covariance function has been shown to be valid on spheres when the Euclidean distance is replaced with the great circle distance. \citep{gneiting2013strictly, huang2011validity}.
We assume $\mathbf{\Sigma}_{\gamma_0}=\sigma^2_{\gamma_0}\mathbf{R}(\psi_{\gamma_0})$, $\mathbf{\Sigma}_{\gamma_1}=\sigma^2_{\gamma_1}\mathbf{R}(\psi_{\gamma_1})$ and $\boldsymbol{\Sigma}_\Delta = \sigma^2_\Delta \mathbf{R}(\psi_\Delta)$,  where $\mathbf{R}(\psi_{\gamma_0})$, $\mathbf{R}(\psi_{\gamma_1})$ and  $\mathbf{R}(\psi_\Delta)$ are also governed by an exponential covariance function with great circle distance. 

\subsection{MCMC Sampling}\label{sec:mcmc}

To describe the details of our sampling, we first write Model (\ref{eq:1}) in a matrix form. 
Let $\mathbf{1}_\tau^+$ be a $MN$-length vector of 0's and 1's, where 1 indicates that the corresponding time index in $Y(\bs_i,t_j)$ is greater than $\tau(\bs_i).$
Let $\mathbf{1}_\tau^-=\mathbf{1}_{MN} - \mathbf{1}_\tau^+$ be the vector of pre-changepoint indices. Then, Model (\ref{eq:1}) can be written as 
\begin{align} \label{eq:mat}
\mathbf{Y} &= \boldsymbol{\mu}_\tau + \mathbf{U} + \boldsymbol{\epsilon}_1\circ \mathbf{1}_\tau^- + \boldsymbol{\epsilon}_2 \circ \mathbf{1}_\tau^+,\\
\shortintertext{where }
\boldsymbol{\mu}_\tau &= \boldsymbol{\mu}_1\circ \mathbf{1}_\tau^- + \boldsymbol{\mu}_2\circ \mathbf{1}_\tau^+,
\end{align}
where $\circ$ denotes the Schur product operator, and $\boldsymbol{\mu}_1, \boldsymbol{\mu}_2, \boldsymbol{\epsilon}_1, \boldsymbol{\epsilon}_2$ denote the $MN$-length vector of $\mu_1(\bs_i, t_j)$, $\mu_2(\bs_i, t_j)$, $\epsilon_1(\bs_i, t_j)$ and $\epsilon_2(\bs_i, t_j)$ for $i=1,\ldots, N$ and $j=1,\ldots, M$. Let $\boldsymbol{\tau}$ be the vector of $\begin{pmatrix} \tau(\bs_1),\ldots, \tau(\bs_N)^T \end{pmatrix}$. The full conditional likelihood of $\bf Y$ given $\mathbf{U}, \boldsymbol{\tau}, \boldsymbol{\mu}_1, \boldsymbol{\mu}_2, \sigma_1^2,\sigma_2^2$ is

\begin{align} \label{eq:lik1}
f(\mathbf{Y}|\cdot)\propto |{\bf\Sigma}_Y|^{-1/2} \exp\left(- \frac{1}{2\sigma_1^2\sigma_2^2}(\mathbf{Y} - \boldsymbol{\mu}_\tau - \mathbf{U})^T{\bf\Sigma}_Y^{-1}(\mathbf{Y} - \boldsymbol{\mu}_\tau - \mathbf{U}) \right),
\end{align}
where ${\bf\Sigma}_Y = \sigma_2^2\text{diag}(\mathbf{1}_\tau^-)+\sigma_1^2\text{diag}(\mathbf{1}_\tau^+)$ is a diagonal matrix of pre and post-changepoint measurement variances.  Alternatively, integrating out $\mathbf{U}$ gives us 
\begin{align} \label{eq:lik2}
f(\mathbf{Y}|\cdot)\propto |{\bf\Sigma}_Y|^{-1/2} \exp\left(- \frac{1}{2\sigma_1^2\sigma_2^2}(\mathbf{Y} - \boldsymbol{\mu}_\tau)^T({\bf\Sigma}_Y+ {\bf\Sigma}_U)^{-1}(\mathbf{Y} - \boldsymbol{\mu}_\tau) \right).
\end{align}

There are two main challenges in MCMC sampling for our model. First, Section 1 of the Supplement \citep{supp} shows that obtaining the posterior for $\mathbf{U}$ involves inverting the covariance matrix ${\bf\Sigma}_Y^{-1} + {\bf\Sigma}_U^{-1}$.
If ${\bf\Sigma}_Y$ is a constant diagonal matrix, then because $\bf\Sigma_U$ is separable we can sample $\mathbf{U}$ using eigendecompositions of $M\times M$ and $N \times N$ matrices instead of having to factorize a $MN\times MN$ covariance matrix \citep{stegle11}. However, (\ref{eq:lik1}) shows that the matrix ${\bf\Sigma}_Y$ involves vectors $\mathbf{1}_\tau^-$ and $\mathbf{1}_\tau^+$ which depends on $\tau(\bs)$, and thus is not a constant diagonal matrix. As a consequence, we have to invert a  $MN\times MN$ covariance matrix. This significantly amplifies the computational cost associated with sampling $\mathbf{U}$.
Secondly, sampling both $\mathbf{U}$ and $\boldsymbol{\mu}_\tau$ can in practice lead to an identifiability issue, since both $\mathbf{U}$ and $\boldsymbol{\mu}_\tau$ have the potential to capture the trend in the data. While integrating out $\mathbf{U}$ from the model could solve these problems, retaining $\mathbf{U}$ is more desirable from a computational perspective. This stems from the fact that $\mathbf{Y}$ is no longer conditionally independent when the conditioning on $\mathbf{U}$ is removed, as indicated by the non-diagonal covariance matrix ${\bf\Sigma}_Y + {\bf\Sigma}_U$ in (\ref{eq:lik2}). Consequently, sampling other parameters whose full conditionals involve the likelihood of $\mathbf{Y}$ becomes significantly more intricate. 

To address the first issue, we propose a simple but effective approach. We assume, without the loss of generality, that the variance increases after the changepoint and write the post-changepoint measurement error as $\boldsymbol{\epsilon}_2 = \boldsymbol{\epsilon}_1 + \boldsymbol{\epsilon}_\gamma, $ where $\boldsymbol{\epsilon}_\gamma \iidsim N(0,\sigma^2_\gamma)$ is a vector of white noise independent of $\boldsymbol{\epsilon}_1.$  Then, we can rewrite
 $f(\boldsymbol{Y}\lvert \cdot)$ in (\ref{eq:lik1}) as 
\begin{align}
f(\boldsymbol{Y}\lvert \cdot)\propto \sigma_1^{MN}\exp\left( -\frac{1}{2\sigma_1^2}(\mathbf{Y} - \boldsymbol{\mu_\tau} - \mathbf{U} - \boldsymbol{\epsilon_\gamma}\circ \mathbf{1_\tau^+})^T(\mathbf{Y} - \boldsymbol{\mu_\tau} - \mathbf{U} - \boldsymbol{\epsilon_\gamma}\circ \mathbf{1_\tau^+}) \right) \label{eq:lik3}
\end{align}
and add
$\boldsymbol{\epsilon_\gamma} \sim N(\mathbf{0}, \sigma^2_\gamma)$
to the level 2 prior. 
The full conditional for $\mathbf{U}$ now involves the covariance matrix $(\sigma^2_1 \mathbf{I}_{MN})^{-1} + \bf\Sigma_U^{-1},$ which can be factored into a Kronecker product of a spatial and a temporal covariance matrix. 
If the variance decreases after the changepoint, 
we simply let $\boldsymbol{\epsilon}_1 = \boldsymbol{\epsilon}_2 + \boldsymbol{\epsilon}_\gamma$ where $\boldsymbol{\epsilon}_\gamma$ now is independent with $\boldsymbol{\epsilon}_2$. Then, we replace $\boldsymbol{\epsilon_\gamma} \circ \mathbf{1_\tau^+}$ with $\boldsymbol{\epsilon_\gamma} \circ \mathbf{1_\tau^-}$ and $\sigma^2_1$ with $\sigma^2_2$ in (\ref{eq:lik3}).  One caveat of this approach is that we need to pre-specify whether the variance increases or decreases beforehand. We recommend trying both scenarios in parallel and choosing a better model based on Deviance Information Criterion (DIC).

\begin{figure}
\centering
\begin{tikzpicture}[node distance=0cm and 2cm]
\node(0) [title,draw=none] {\textbf{Ordinary Gibbs Sampler}}; 
\node (1) [block,align=left,below of = 0, yshift=-1.5cm] {Step 1: $\tau \sim p(\tau| Y,U',\theta',\alpha')$\\ 
Step 2: $\theta \sim p(\theta|Y,U',\alpha',\tau)$\\ 
Step 3: $\alpha \sim p(\alpha| Y,U',\theta,\tau)$\\ 
Step 4: $U \sim p(U|Y,\alpha,\tau,\theta)$};
\node (4) [block,align=left, right of = 1, xshift=8cm] {Step 1: $\tau \sim p(\tau| Y,\theta',\alpha')$\\ Step 2: $\alpha \sim p(\alpha| Y,\theta',\tau)$\\ Step 3: $U \sim p(U|Y,\tau,\alpha,\theta')$ \\ Step 4: $\theta \sim p(\theta|Y,U,\tau,\alpha)$};
\node (5) [title,draw=none,above of = 4,yshift=1.5cm]{\textbf{Partially Collapsed Gibbs Sampler}};

\draw [arrow] (1) -- node[anchor=west]{}(4);
\end{tikzpicture}
\caption{Sampling procedure for Partially Collapsed Gibbs (PCG) sampler. In our context, $p(\cdot)$ denotes the target posterior distribution and $\theta = (\sigma^2_1,\sigma^2_2,\sigma^2_\Delta, \boldsymbol{\epsilon_\gamma},\psi_U,\phi_U,\psi_\Delta,\psi_{\gamma_0},\psi_{\gamma_1},\boldsymbol{\beta})^T.$ Compared to the ordinary Gibbs sampler, the PCG sampler drops the conditioning on $\mathbf{U}$ when sampling $\tau$ and $\alpha$, and then modifies the order of sampling in the ordinary Gibbs to ensure the stationary distribution of the chain under PCG.} 
\label{pcgs}
\end{figure}
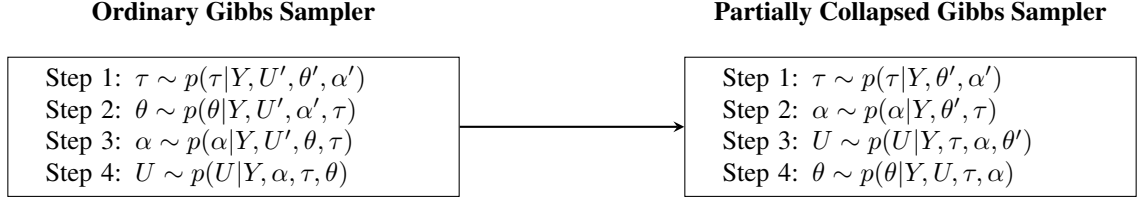

To address the potential challenge of identifiability between $\bf U$ and $\boldsymbol{\mu}_\tau$ in sampling, we adopt the Partially Collapsed Gibbs (PCG) sampler proposed by \cite{van2008partially}, which allows us to remove the conditioning on $\mathbf{U}$ when sampling $\boldsymbol{\tau}$ and $\boldsymbol{\alpha} = (\alpha_0,\gamma_0,\gamma_1)^T$. We refer to \cite{van2008partially} and \cite{van2015metropolis} for more details on how to derive the PCG sampler from an ordinary Gibbs sampler. Figure \ref{pcgs} illustrates the sampling procedure for our model under the PCG sampler. It is important to note that, unlike ordinary Gibbs samplers, the order of draws in the PCG sampler must be maintained, as permuting the order may alter the stationary distribution of the chain in the PCG sampler \citep{van2015metropolis}. 

The following discusses the sampling for each parameter in turn. 
The full conditionals for $\mathbf{U}$ and $\boldsymbol{\epsilon_\gamma}$ are $MN$-dimensional multivariate normal distributions. To sample $\mathbf{U}$, we leverage the compatibility of a Kronecker product plus a constant diagonal matrix with the eigenvalue decomposition \citep{stegle11}, which allows us to sample $\mathbf{U}$ using eigendecompositions of $M\times M$ and $N \times N$ matrices. While the posterior for $\boldsymbol{\epsilon_\gamma}$ is no longer identically distributed, Section 1 of the Supplement \citep{supp} shows that its independence structure is still preserved. Thus, posterior samples can be drawn easily even when the spatial and/or temporal dimension is high. See the Supplement Material \citep{supp} for more details.

Parameters $\boldsymbol{\Delta}, \boldsymbol{\gamma_{0R}}, $ and $\boldsymbol{\gamma_{1R}}$ are sampled via a Metropolis Hastings algorithm. While the full conditionals for $\boldsymbol{\gamma_{0R}}$ and $\boldsymbol{\gamma_{1R}}$ have closed forms, the resulting distribution involves $MN \times MN$ covariance matrices, making the sampling computationally expensive. To achieve efficient mixing, we employ an adaptive component-wise Metropolis-within-Gibbs \citep{roberts09} to propose a new state for each of the $N$ locations separately and automatically tune the acceptance ratios for each location as closely as possible to 0.44, a ratio considered optimal for one-dimensional Gaussian proposals \citep{roberts2001optimal}. 

Finally, the parameters $\sigma^2_U, \sigma_1^2, \sigma_2^2, \sigma^2_\Delta, \sigma^2_{\gamma_0}, \sigma^2_{\gamma_1}, \boldsymbol{\beta},$ and $\boldsymbol{\alpha} = (\alpha_0, \gamma_0, \gamma_1)^T$ are assigned conjugate priors and estimated via a Gibbs sampler. We specify weakly independent normal prior for $\boldsymbol{\alpha}$ and $\boldsymbol{\beta}$ and inverse gamma priors for variance parameters $\sigma^2_1, \sigma^2_2, \sigma^2_U,$ and $\sigma^2_\Delta.$ For $\sigma^2_{\gamma_0}$ and $\sigma^2_{\gamma_1},$ we use a truncated inverse gamma prior and choose the upper truncation points to be a small fraction of $\gamma_{0F}$ and $\gamma_{1F}.$ This encourages the magnitude of random effect to be smaller than that of the fixed effect, which further ensures identifiability. The parameters $\psi_U,\phi_U,\psi_\Delta, \psi_{\gamma_0}, \psi_{\gamma_1}$ are given uniform priors and are sampled via the Metropolis Hastings algorithm. For $\tau_0,$ any discrete probability distribution that takes values in $\{1,2, \ldots, M\}$ can be used as a prior. The full conditional distributions and sampling details are outlined in Section 1 of the Supplement \citep{supp}.

\section{Simulation Study}\label{sec:simstudy}

We perform simulation studies to evaluate the accuracy of our model in detecting and estimating three different types of changes -- mean shift only, variance shift only, and mean and variance shift. We compare our method with a 
univariate time series changepoint detection method derived by only considering $N=1$ in our model. 
This univariate method still considers temporal correlation but no spatial correlation and is applied to each of the spatial locations separately. We also explore how the magnitude of change affects the model's performance. 

\subsection{Setup}
To generate data for simulation, we select $N = 121$ spatial locations on an $11\times 11$ grid and randomly choose one of the grid locations as $\mathbf{s_0}$. The grid is on a surface of a sphere with latitudes from $0^\circ$ to $50^\circ$ and longitudes from $0^\circ$ to $50^\circ,$ both in an increment of $5^\circ$. At each location, we consider a time series of length $M = 61$ and generate $\tau(\mathbf{s})$ using (\ref{eq:cp}), with $\tau_0 = 18, \boldsymbol{\beta} =  (1.5, 1)^T,$ $\sigma^2_\Delta = 1,$ and $\psi_\Delta = 0.5$. These values were chosen such that the number of locations with changepoint $|\{\bs : \tau(\mathbf{s}) < M\}|$ is comparable to the number of locations without changepoints $|\{\bs : \tau(\mathbf{s})\ge M\}|$ on average. Since a changepoint detection is equivalent to a classification problem, having a balance between the two classes ensures our evaluation focuses on method performance without being affected by class imbalance. 

We use model (\ref{eq:1}) to generate data but replace $\mu_2(\bs, t)$ in model (\ref{eq:RE2}) by $\mu_2(\bs, t)=\alpha_0+\gamma_0$. This simple version of $\mu_2(\bs, t)$ serves the purpose of changepoint detection evaluation and offers a fair comparison with the classic method for which the default model contains no slope after the changepoint.  
Without loss of generality, we fix the pre-changepoint mean $\alpha_0=0$ and variance $\sigma^2_1=1$. We let $\gamma_0$ take values in $\{0,1.5,2,3\}$ and $\sigma^2_\gamma$ in $\{0,3,5\}$ to study the model performance under different strengths of mean and variance shift. The parameters $\sigma^2_U,$ $\phi_U,$ and $\psi_U$ are set to 1, 1.5, and 2, respectively, to emulate spatial and temporal dependence observed in real data based on our exploratory data analysis of global surface temperature data. We run 100 simulations for each setting. The ratio of the number of locations without changepoints to those with changepoints in our simulated data ranges from 32:89 to 101:20, with an average of 57:64.

In implementing our method, we assume no prior knowledge about $\tau_0$ and use a discrete uniform prior. This is slightly more challenging than our real problem, since for the Pinatubo data, we know $\tau_0$ is likely right after the eruption. We give weakly informative prior for other unknown parameters. We use the posterior mode of $\min\{M,\tau(\bs)\}$ as the estimated changepoint for location $\bs$. If the mode for $\min\{M,\tau(\bs)\}$ is at $M$, we determine that there is no changepoint at $\bs$.

To evaluate the performance of our method, we use the metrics of false positive rate (FPR) and false negative rate (FNR) to measure the accuracy of the changepoint detection, and use root mean squared error (RMSE),  empirical coverage of the 95\% credible interval (CI), and length of the CI to assess the changepoint estimation. FPR is defined as the ratio of the number of falsely detected locations to the total number of locations without changepoints, and FNR is the ratio of the number of falsely undetected locations to the total number of locations with changepoints. When calculating RMSE, we treat the true changepoint as having value $M$ (i.e., $\tau(\mathbf{s})=M$) when there is no changepoint at location $\bs$. A combination of a narrower credible interval with empirical coverage closer to the nominal level indicates a more precise uncertainty quantification. The formula for FPR, FNR, and RMSE are given below, where $\tau({\bf s})$ denotes the true changepoint for location ${\bf s}$, $\hat{\tau}({\bf s})$ denotes the detected changepoint, and $\lvert A \rvert$ is the cardinality of set A:
\begin{align*}
    FPR &= \frac{FP}{FP+TN} = \frac{\lvert \{ {\bf s} : \hat{\tau}({\bf s}) < M, \tau({\bf s}) = M \} \rvert }{ \lvert \{ {\bf s} : \tau({\bf s}) = M \} \rvert }\\
    FNR &= \frac{FN}{FN+TP} = \frac{\lvert \{ {\bf s} : \hat{\tau}({\bf s}) = M, \tau({\bf s}) < M \} \rvert }{ \lvert \{ {\bf s} : \tau({\bf s}) < M \} \rvert }\\
    RMSE &= \sqrt{\frac{ \sum_{\bf s} (\tau({\bf s}) - \hat{\tau}({\bf s}))^2 }{N}}
\end{align*}

We compare the FPR, FNR and RMSE of our spatio-temporal model with the univariate method applied to each location (hereinafter 1D method). Through experimentation, we picked a conservatively large iteration number of $20,000$ and a burn-in size of $10,000$ to ensure MCMC convergence based on Gelman-Rubin diagnostic \citep{gelman1992inference}.


\subsection{Results}
\begin{figure}[!ht]
\centering
\includegraphics[width=14.2cm]{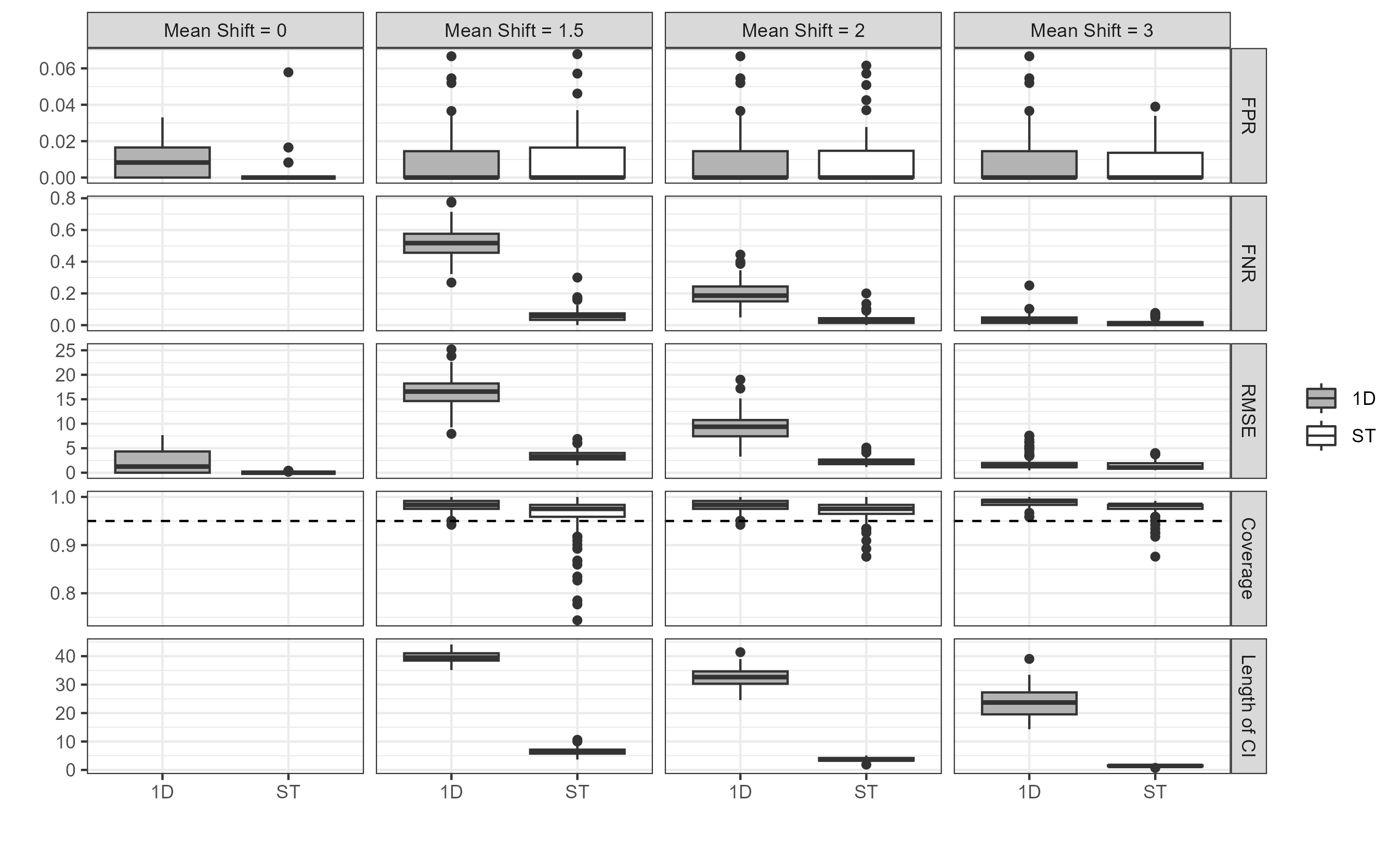}
\caption{\label{fig:sim_mean_only} Boxplots of RMSE, FPR, FNR, the empirical coverage probability  and the length of 95\% credible intervals under different mean shift signal strength. 
“ST” is our spatio-temporal model and “1D” is the univariate method. FNR is not defined under Mean Shift=0.} 
\end{figure}

Figure \ref{fig:sim_mean_only} summarizes the results for detecting and estimating changepoints only in the mean under different magnitudes of change. Mean shift of zero indicates no changepoint, i.e. $\tau(\mathbf{s}) = M$ for all locations, while a larger mean shift indicates a stronger signal. 

The FPR of both methods are comparable when a changepoint is present. Still, our method seems more stable when signal is strong. When there is no changepoint, our method returns a much lower FPR. The FNRs of the two methods are comparable when the signal is strong, but our method largely outperforms the 1D under a weaker signal. Note that FNR is not defined when there is no changepoint. 
Both the 1D and our method achieve smaller RMSE when the signal is stronger, but our method outperforms the 1D method across all signal strengths with significantly smaller RMSE. The advantage of our method in terms of RMSE is more pronounced when the signal is weaker. 
The empirical coverage of the $95\%$ credible intervals obtained from the 1D method appears to be above the nominal level for all tested cases. However, the mean empirical coverage of our method is centered around the nominal level. More importantly, our method results in a much narrower and thus more informative credible interval compared to the 1D method. Unsurprisingly, the length of the credible interval gets wider as the signal becomes weaker.
Although there is no simple interpretation for the empirical coverage of Bayesian credible intervals as there is for confidence intervals, our results still suggest a positive view of our credible intervals. The lengths of the credible interval appear to be informative given that the maximum interval length is $60$. Unsurprisingly, the length of the credible interval gets wider as the signal becomes weaker.

\begin{figure}[!ht]
\centering
\includegraphics[width=14.2cm]{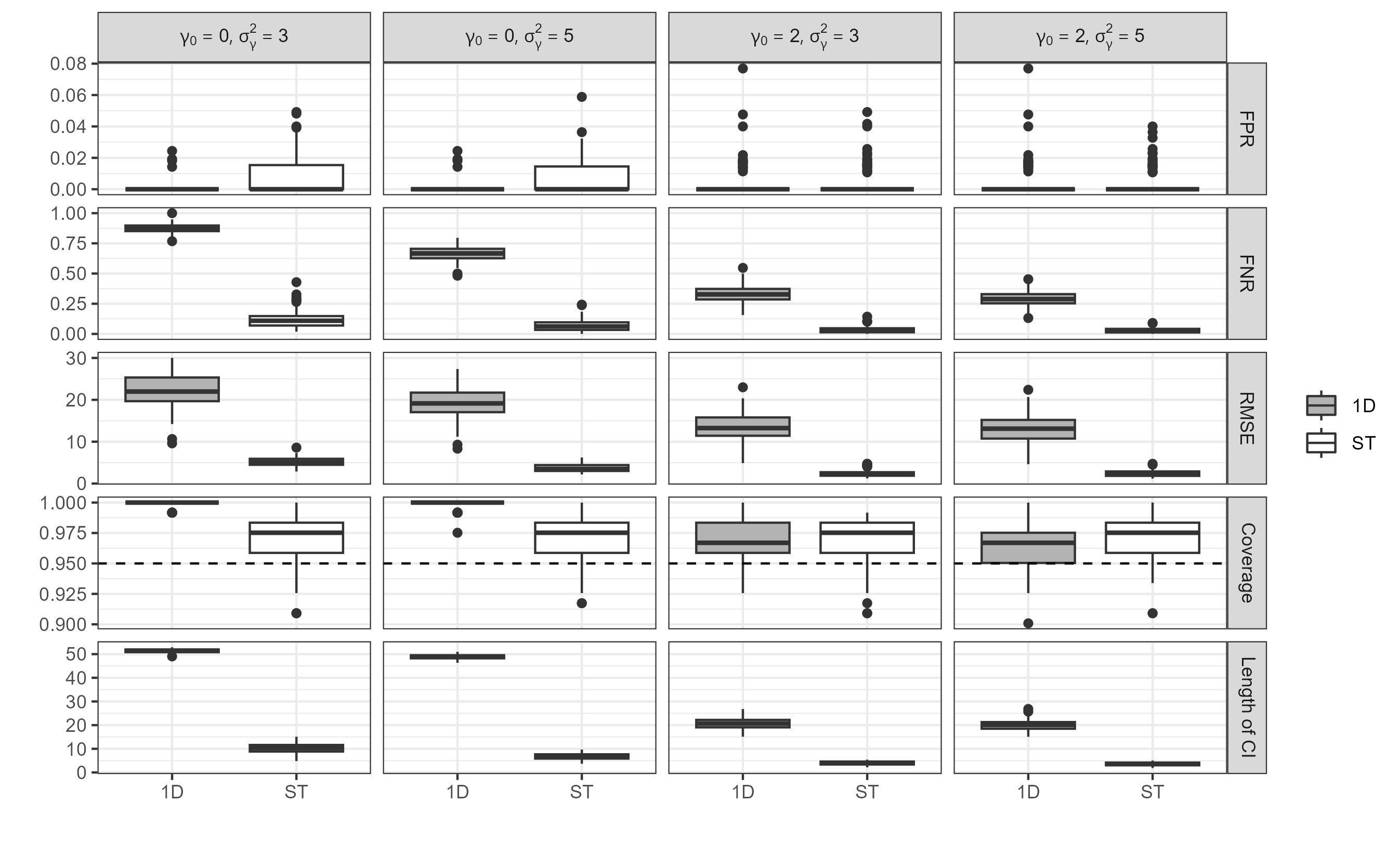}
\caption{\label{fig:sim_meanvar} Boxplots of RMSE, FPR, FNR, the empirical coverage probability  and the length of 95\% credible intervals under four settings comprised of two $\gamma_0$ values representing mean change and two $\sigma^2_\gamma$ values representing variance change. $\gamma_0=0$ indicates no mean change. “ST” is our spatio-temporal model and “1D” is the univariate method. }
\end{figure}

Figure \ref{fig:sim_meanvar} shows the results for both variance shift only and mean and variance shift combined. We show results for $\gamma_0 \in \{0,2\}$ associated with zero and positive mean shift, and the variance difference parameter $\sigma^2_\gamma \in \{3,5\}$, representing two different levels of variance shift. The results for $\gamma_0 \in \{1.5,3\}$ and $\sigma^2_\gamma \in \{3,5\}$ are provided in Section 2 of the Supplement \citep{supp}. 

In the case of variance only shift, our method significantly outperforms the 1D method, with even a wider margin, in all metrics except for FPR. The 1D method appears to be very conservative in detecting variance only shift, resulting in a lower FPR but an extremely high FNR. This implies that the 1D method is less sensitive to variance changes than mean changes. 

For combined mean and variance shift, the results seem to be very similar to those for mean shift only. The FPR is comparable for both methods, while the FNR and RMSE are much lower for our method. In this combined case, the empirical coverage of credible intervals for the 1D method is closer to, though still larger than, the nominal level. Once again, the credible intervals obtained from our method are much shorter than the 1D method without sacrificing their empirical coverage. 

For each simulation, we calculate the posterior estimate of the parameter by taking posterior mode for $\tau_0$ and posterior mean for all other parameters. To give an example of parameter estimation with our method,  we summarize the parameter estimates for the simulations with mean shift of 1.5 and $\sigma^2_\gamma = 0$ in Table \ref{tab:sim_summ}.  The second and third columns of the table report the mean estimates and the $95\%$ credible intervals across the 100 simulations. We observe that the true value is captured by the $95\%$ interval for all parameters in the model.

\begin{table}[h!]
\centering
\begin{tabular}{clcc}
\hline
Parameter & True Value & Mean of the estimate & 95\% CI of the mean \\\hline
$\alpha_0$ & 0 & 0.0014  & (-0.0831, 0.0754) \\
$\gamma_{0F}$ & 1.5 & 1.4616 & (1.3260, 1.6342) \\
$\sigma^2_1$ & 1 & 1.0309 & (0.8844, 1.2461) \\
$\sigma^2_U$ & 1 & 0.9335 & (0.7634, 1.1445) \\
$\psi_U$ & 2 & 1.8901 & (1.5467, 2.3630) \\
$\phi_U$ & 1.5 & 1.4489 & (1.1393, 1.7735)\\
$\tau_0$ & 18 & 16 & (8, 20)\\
$\beta_1$ & 1.5 & 1.4366 & (0.9382, 2.3505) \\
$\beta_2$ & 1 & 1.0408 & (0.6369, 1.6172)\\
$\sigma^2_\Delta$ & 1 & 0.8644 & (0.6362, 1.3412) \\
$\psi_\Delta$ & 0.5 & 0.6572 & (0.3544, 1.0177)\\
\hline
\end{tabular}
\caption{\label{tab:sim_summ} Summary of parameter estimates for a mean shift of 1.5 and variance shift of 0.}
\end{table}

Overall, our method outperforms the 1D method by taking advantage of the spatial correlation in the changepoint process for spatially indexed time series. The importance of borrowing strength from neighboring locations becomes more pronounced and necessary when the change signal is weaker. Similar phenomena were observed by \cite{wang2023asynchronous}. 

\section{Impact of Mt Pinatubo Volcanic Eruption}\label{sec:results}
We apply our proposed method to AOD and surface temperature data to detect changes that are possibly caused by the aerosols injected into the stratosphere by the Mt. Pinatubo eruption in June 1991. All MCMC convergence in the data analyses have been verified by the Gelman-Rubin diagnostic using three parallel chains with different initial values for the parameters.

\subsection{Aerosol Optical Depth}

Following the suggestions in Section \ref{sec:mcmc}, we apply model (\ref{eq:1}) on the aerosol data with both $\sigma^2_1 < \sigma^2_2$ and $\sigma^2_1 > \sigma^2_2$ since we have no prior knowledge about whether the variance increases or decreases after changepoint. In addition, we also test the case of $\sigma^2_1 = \sigma^2_2,$ which means that the change is solely due to a mean shift.

We set $\bs_0$ to be at ($123.4375^\circ$E, $19^\circ$N), the center of the grid cell that contains the location of Mt. Pinatubo ($120^\circ$E, $15^\circ$N).  We anticipate $\bs_0$ has the earliest changepoints among all grid points.
We use a discrete uniform prior for $\tau_0$ and weakly informative priors for all other parameters. 
Table \ref{tab:totext_summ1} compares the DIC value and the posterior mean of key parameter estimates for the three forms of variance shift models.
We observe that the equal variance model achieves the lowest DIC value. Consistent with this finding, the estimated variance shift parameter $\sigma^2_\gamma$ for the variance increase or decrease assumptions is very small compared to $\sigma^2_1$ or $\sigma^2_2$. Thus, we proceed with our inference based on the assumption $\sigma_1^2=\sigma_2^2$, that is, the change is only due to a mean shift. 
 \begin{table}
\centering
  \begin{tabular}{c|ccc |c} 
  \hline
    $\phantom{d}$ & $\sigma^2_1 = \sigma^2_2$ & $\sigma^2_1 < \sigma^2_2$ & $\sigma^2_1 > \sigma^2_2$ & 95 \% CI\\ 
    DIC & 3,526,069 & 3,638,721 & 3,719,177 & for $\sigma^2_1 = \sigma^2_2$\\
\hline

  $\alpha_0$ & 0.2754 & 0.1913 & 0.2499 & (0.0754,0.4808)\\
  $\gamma_{0F}$ & 4.2408 & 4.2825 & 4.923 & (3.4503,4.7342)\\ 
  $\gamma_{1F}$ & -0.0820 & -0.0888 & -0.0881 & (-0.0996,-0.0646) \\
  $\sigma^2_U$ & 1.1449 & 1.1391 & 1.1410 & (1.0131,1.3127) \\
  $\sigma^2_1$ & 0.3512 & 0.2790 & 0.1409 & (0.2563,0.5245)\\
  $\sigma^2_2$ & NA & 0.3092 & 0.1348 & NA\\
  $\sigma^2_\gamma$ & NA & 0.0302 & 0.0061 & NA\\
  $\beta_1$ & -0.0186 & -0.0092 & -0.0151 & (-0.1488,0.1020) \\
  $\beta_2$ & 0.5404 & 0.4447 & 0.4490 & (0.1562,0.9623) \\
  $\sigma^2_\Delta$ & 0.9036 & 0.8288 & 0.8587 & (0.4766,1.786)\\
  $\psi_U$ & 2.5679 & 2.8244 & 2.6175 & (2.0745,2.9091)\\
  $\phi_U$ & 0.8336 & 0.8965 & 0.8229 & (0.7161,0.9198)\\
  $\psi_\Delta$ & 0.8084 & 0.8489 & 0.8212 & (0.3817,1.7390)\\
  $\sigma^2_{\gamma_0}$ & 1.5794 & 1.3702 & 1.7418 & (1.0757,2.1393)\\
  $\sigma^2_{\gamma_1}$ & 0.0006 & 0.0007 & 0.0007 & (0.0004,0.0011)\\
  $\psi_{\gamma_0}$ & 6.7958 & 7.5673 & 5.8663 & (4.1403,10.7769)\\ 
  $\psi_{\gamma_1}$ & 6.5380 & 5.9436 & 5.8552 & (4.3915,9.9309)\\
  \hline
  \end{tabular}
\caption{\label{tab:totext_summ1} DIC values and posterior summary of parameters under different model assumptions for AOD data. The credible intervals in the last column are derived under $\sigma^2_1 = \sigma^2_2$.}
\end{table}

We report the 95\% credible intervals of key parameters under the assumption of $\sigma_1^2=\sigma_2^2$ in Table~\ref{tab:totext_summ1}. The significantly positive estimate of $\gamma_{0F}$ indicates an elevated aerosol level immdediately after the volcanic eruption, and the negative estimate of $\gamma_{1F}$ indicates the subsequent restoration of aerosol level post-eruption, consistent with the anticipated trend. The diffusion parameter for longitude, $\beta_1$, is not significantly different from 0. In contrast, there is strong evidence to show that the diffusion parameter for latitude, $\beta_2$, is substantially greater than 0. This observation aligns with the spatial pattern of changepoints predominantly driven by latitude, as depicted in Figure \ref{fig:heatmap_eda}(a). 

\begin{figure}[!ht]
\centering
  \includegraphics[width=14.2cm]{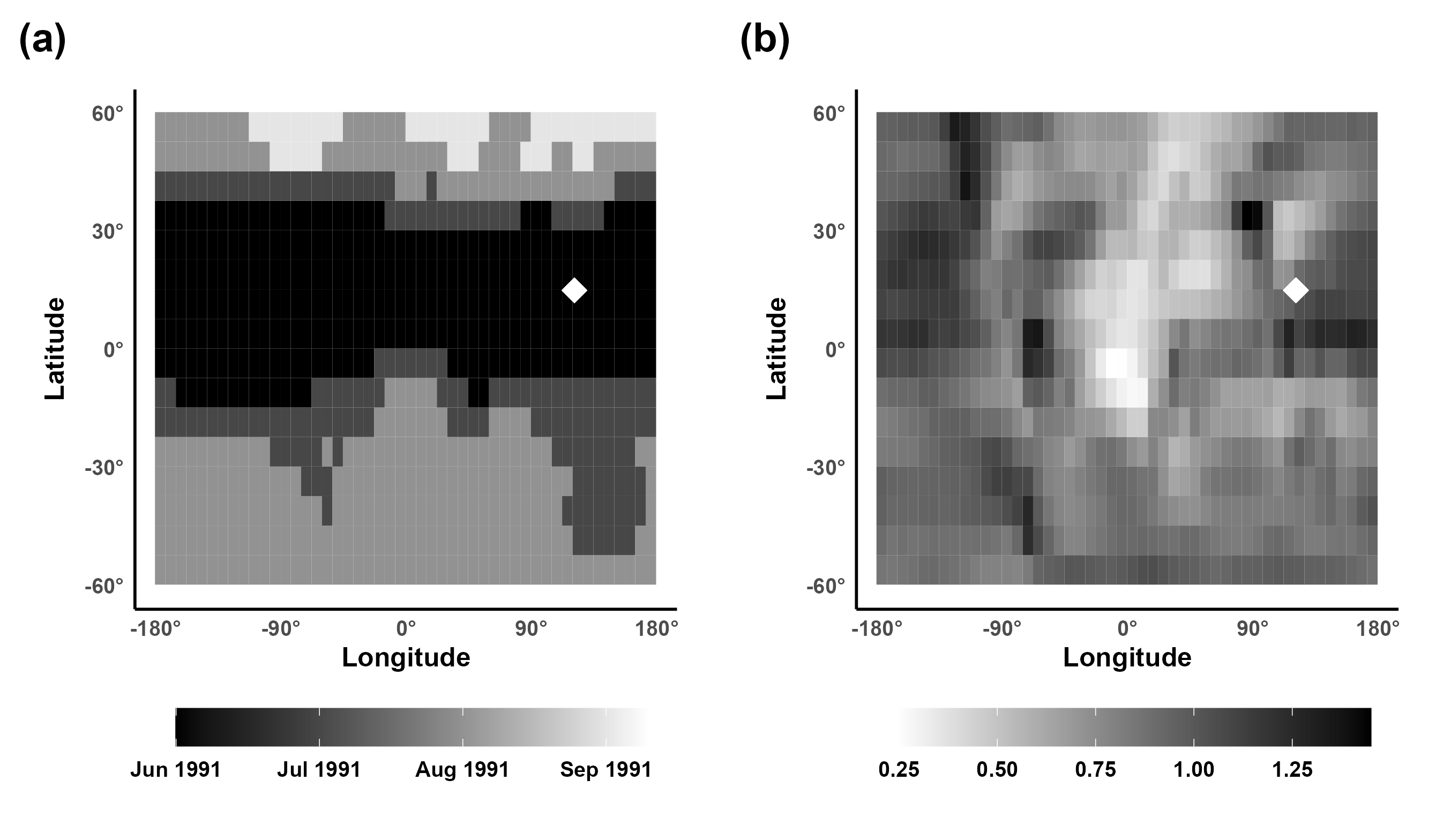}%
  \caption{\label{fig:frog} (a) Heatmap of detected changepoints for AOD. Darker color indicates earlier change. (b) Heatmap of estimated change amount. Darker color indicates larger magnitude of change. The white diamond marks the location of Mt. Pinatubo. }
\end{figure}

We determine the estimated changepoints by taking the posterior mode of the floor function of $\min\{M,\tau(\mathbf{s})\}$ at each location. Figure \ref{fig:frog}(a) shows the heatmap of estimated changepoints. Our model detects a changepoint at all locations, with estimated values ranging from Jun 1991 to Sep 1991. The estimated changepoints show a pattern driven more by latitude than longitude, with the earliest changepoints predominantly occurring along latitudes $3.5^\circ$S through $34^\circ$N. This pattern aligns with our exploratory data analysis and is consistent with existing literature, which reports that the Pinatubo aerosol layer circled the Earth in 21 days and had spread to latitudes around $30^\circ N$ and $10 ^\circ S$ in the same period \citep{self1993}. \cite{mccormick1992sage} and \cite{stowe1992monitoring} also found the Pinatubo aerosols straddled the equator. 
Figure \ref{fig:frog}(b) displays the estimated mean shift parameter, $\gamma_0(\bs)$, which ranges from $+0.25$ to $+1.5$. The spatial pattern of the estimated mean change matches with that shown in Figure \ref{fig:heatmap_eda}(b), indicating the successful capture of the spatial variability in mean change by our method. 

\begin{figure}[!ht]
\centering
\includegraphics[width=14.2cm]{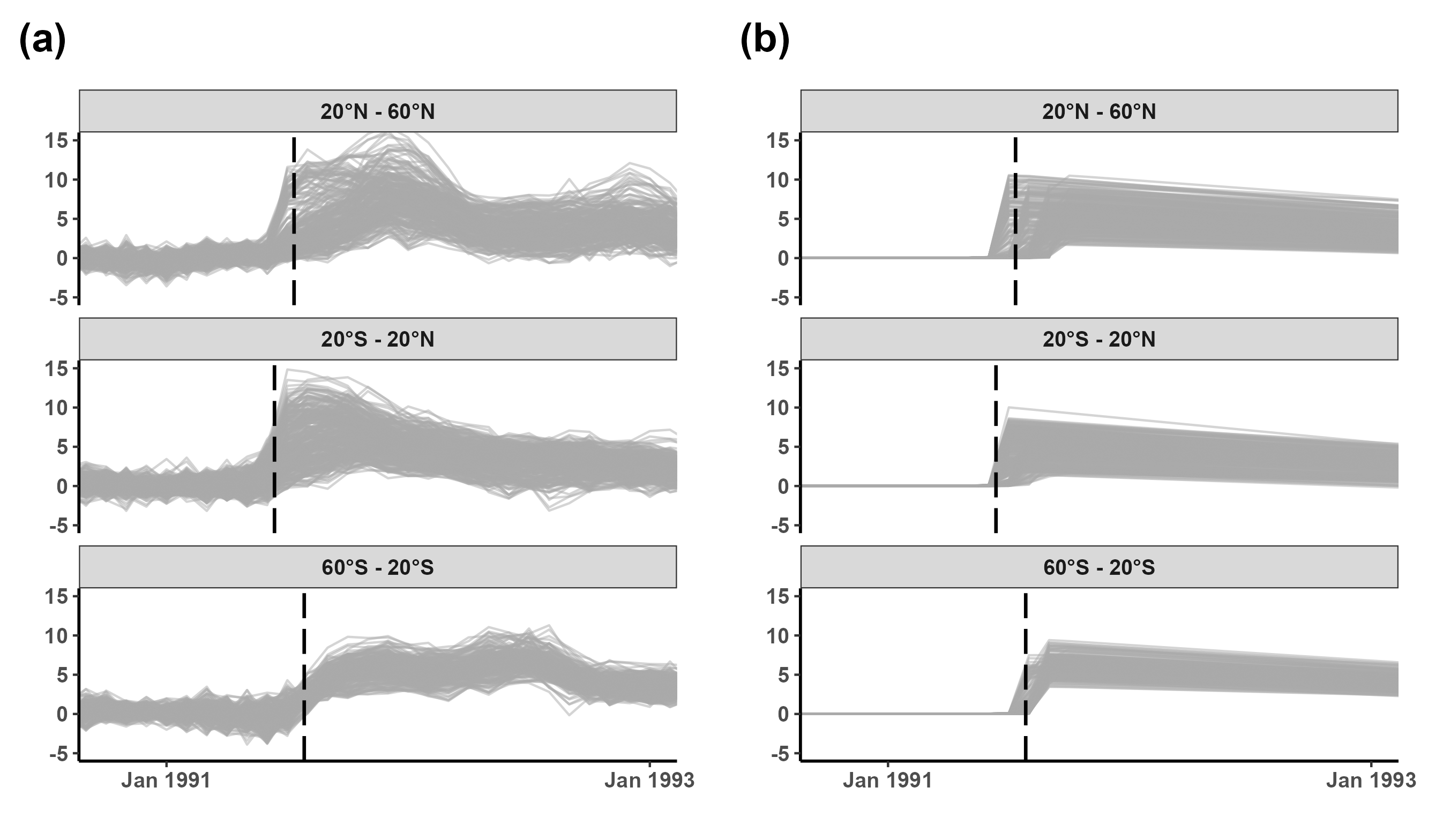}
\caption{\label{fig:totext_mean} Time series of (a) AOD after preprocessing and (b) posterior mean of $\mu_1$ and $\mu_2(\mathbf{s},t)$, zoomed in to years $1991-1993.$ The dashed line marks the average value of the changepoints in latitude bands $(60^\circ S-20^\circ S), (20^\circ S-20^\circ N),$ and $(20^\circ N-60^\circ N)$.} 
\end{figure}

The time series of AOD after preprocessing and the estimated $\mu_1$ and $\mu_2(\mathbf{s},t)$ from our model are shown in Figure \ref{fig:totext_mean}. Our mean estimates follow the same pattern as the preprocessed data, of which the time series close to the equator in the latitude bands $20^\circ S - 20^\circ N$ jump first and tend to have a higher peak compared to those around the latitudes further north or south. These results, together with Figure \ref{fig:frog}, demonstrate that our model effectively captures the diffusion of the impact of the volcanic eruption on AOD. Specifically, the estimated changepoints for locations near the latitude of Mt. Pinatubo ($15^\circ$N) coincide with the month of the eruption, while the changepoints for locations further north or south occur several months later. We also plot the residuals for the AOD series by location in Section 2 of the Supplement \citep{supp}, which verifies that $\sigma_1^2=\sigma_2^2$ is a reasonable assumption. 

\subsection{Zonal surface temperature data}

Unlike the clear signal of volcanic impact on stratospheric AOD data, the impact on surface temperature is visually indiscernible, as observed in Figure \ref{fig:raw_data}. Indeed, the impact was too subtle to be detected when we applied our method to the gridded temperature data as was done for AOD data. Due to the strong latitudinal, or zonal, trend of AOD, the impact of volcanic eruption on surface temperature is not as local as the observed impact on AOD. Changes in surface temperature following the eruption are more noticeable when aggregated by zonal bands, as mentioned in Section \ref{sec:data}. Consequently, we use zonal mean temperatures to trace the impact of the eruption. The mechanism behind how aerosols ejected into the lower stratosphere affect surface temperature is a complex process. Unlike AOD, we do not expect the volcanic impact on surface temperatures to diffuse from the event location \citep{robock1983circumglobal}. Thus, we negate the influence of the distance from $\mathbf{s_0}$ on $\tau_0(\bs)$ by setting
$\boldsymbol{\beta} = 0$ in the mean of $\log(\boldsymbol{\Delta})$. For this data, $\bs_0$ is interpreted as the location corresponding to the smallest $\tau(\bs)$ estimate.  

Previous literature has found that the effect of the Mt. Pinatubo eruption on surface temperature mostly took place in the two-year period following the eruption \citep{robock1983circumglobal, self1993}. Therefore, we give $\tau_0$ an equal weight discrete uniform prior covering the months in the following two years and the last month in the data: $\{1991/06, 1991/07, \ldots 1993/06, 1995/12\}$. The last time point, $1995/12$, indicates the possibility that there is no observed changepoint at any of the locations. 

Similar to the AOD data, we apply Model (\ref{eq:1}) with $\sigma^2_1 < \sigma^2_2$,  $\sigma^2_1 > \sigma^2_2$, and $\sigma^2_1 = \sigma^2_2$ to surface temperature and compare the results of the three models in Table \ref{tab:temp_summ1}. Again, the mean shift only model ($\sigma^2_1 = \sigma^2_2$) achieves the lowest DIC, and the estimated variance change parameter $\sigma^2_\gamma$ for the two variance shift models is very small compared to $\sigma^2_1$ or $\sigma^2_2$. We thus proceed with our analysis under the assumption $\sigma^2_1 = \sigma^2_2$. Table \ref{tab:temp_summ1} reports the posterior mean and the 95\% credible interval of key parameters. Unlike the AOD data, the mean shift parameter $\gamma_{0F}$ for temperature is significantly negative, indicating the global cooling effect of eruption consistent with other literature. The slope parameter $\gamma_{1F}$ is also significantly negative, indicating continued cooling of the global temperature after the changepoint during the time period considered in our data.

 \begin{table}
\centering
  \begin{tabular}{c|ccc | c} 
     \hline
    $\phantom{d}$ & $\sigma^2_1 = \sigma^2_2$ & $\sigma^2_1 < \sigma^2_2$ & $\sigma^2_1 > \sigma^2_2$ & 95\% CI\\ 
    DIC & -1,692 & -1,543 & -1,631 &  for $\sigma^2_1 = \sigma^2_2$\\
    \hline
  $\alpha_0$ & 0.2092 & 0.2117 & 0.2103 & (0.1182,0.3074)\\
  $\gamma_{0F}$ & -0.8440 & -0.8592 & -0.8430 & (-1.1714,-0.5410)\\ 
  $\gamma_{1F}$ & -0.0117 & -0.0116 & -0.0116 & (-0.0213,-0.0028) \\
  $\sigma^2_U$ & 0.7918 & 0.7933 & 0.7901 & (0.7242,0.8655)\\
  $\sigma^2_1$ & 0.1029 & 0.1046 & 0.1068 & (0.0798,0.1436)\\
  $\sigma^2_2$ & NA & 0.1127 & 0.1011 & NA \\
  $\sigma^2_\gamma$ & NA & 0.0081 & 0.0057 & NA\\
  $\sigma^2_\Delta$ & 20.16 & 20.92 & 20.43 & (8.33,56.65)\\
  $\psi_U$ & 4.9675 & 4.8348 & 4.8779 & (4.2658,5.6670) \\
  $\phi_U$ & 1.2129 & 1.2063 & 1.1958 & (1.0662,1.3851)\\
  $\psi_\Delta$ & 1.7805 & 2.0915 & 1.8195 & (0.8843,3.0968)\\
  $\sigma^2_{\gamma_0}$ & 0.0390 & 0.03870 & 0.0312 & (0.0145,0.1516)\\
  $\sigma^2_{\gamma_1}$ & 3.49e-5 & 3.63e-5 & 3.56e-5 & (1.93e-6,9.88e-5)\\
  $\psi_{\gamma_0}$ & 21.46 & 20.63 & 23.38 & (15.29,28.47)\\ 
  $\psi_{\gamma_1}$ & 27.48 & 25.75 & 25.51 & (18.50,28.66)\\

  \hline
  \end{tabular}
\caption{\label{tab:temp_summ1} DIC values and posterior summary of parameters under different model assumptions for surface temperature data. The credible intervals in the last column are derived under $\sigma^2_1 = \sigma^2_2$. }
\end{table}

\begin{figure}
\centering
\includegraphics[width=14.2cm] 
{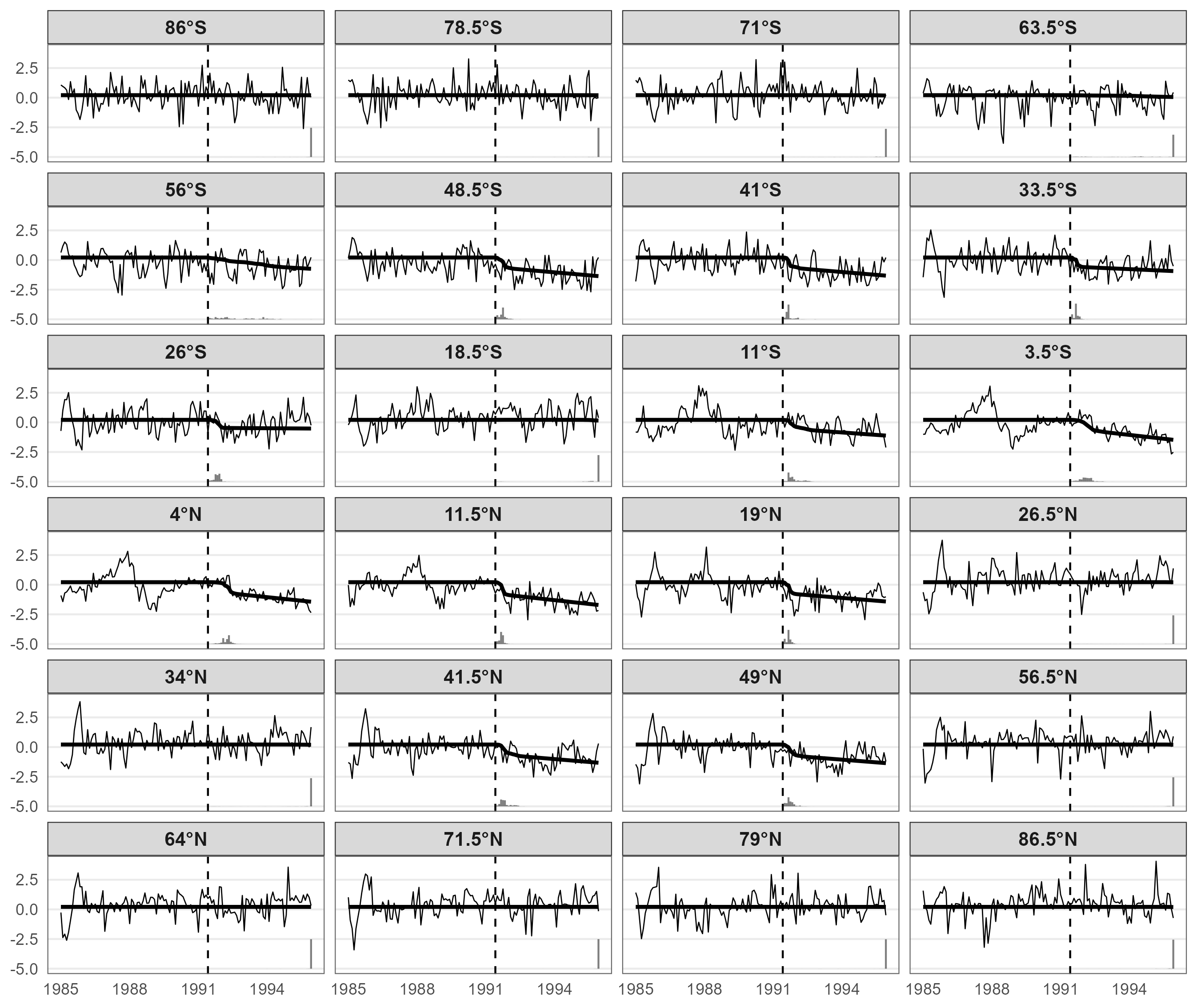}
\caption[Image]{Posterior mean of $\mu_1$ and $\mu_2(\bs, t)$ for surface temperature, separated at the posterior mode of $\tau(\bs)$ (thick black lines). The thin black lines are latitudinal mean temperature series after preprocessing. The posterior distribution of the changepoints are overlayed at the bottom of each plot. 
No changepoints were detected at $86.0^\circ$S - $63.5^\circ$S, $18.5^\circ$S, $26.5^\circ$N, $34.0^\circ$N, and $56.5^\circ$N - $86.5^\circ$N. The dashed vertical lines mark the month of the eruption (1991/06)}
\label{fig:surface_lat_mean}
\end{figure}

Figure~\ref{fig:surface_lat_mean} shows the posterior mean of 
$\mu_1$ and $\mu_2(\bs, t)$ for the temperature series at each latitude, separated at the posterior mode of $\tau(\bs).$
Our method detects changepoints for all latitudes in the range $(56^\circ S - 49^\circ N)$ except for latitudes $18.5^\circ S, 26.5^\circ N, $ and $34.0^\circ N.$ No changepoints were detected in the southern $(86^\circ S - 63.5^\circ S)$ and northern $(56.5^\circ N - 86.5^\circ N)$ ends of the globe. The posterior probability of having no changepoint ranges from 0 to 0.0112 for the detected latitudes and 0.7955 to 0.9986 for the undetected latitudes. The detected changepoints range from Sep 1991 - May 1992. See Table 1 in the Supplement \citep{supp} for further details. 
The lack of changepoint in the latitudes $18.5^\circ S, 26.5^\circ N, 34.0^\circ N$ may potentially be due to low signal-to-noise ratio in the subtropics (roughly 23$^\circ$ to 35$^\circ$ north and south) which are particulary synoptically active \citep{mcclain2004subtropical, ryoo2008variability, lensky2018synoptic}. As shown in Figure \ref{fig:surface_lat_mean}, at all these three latitudes,  we observe a noticeable drop in the temperature shortly after the Pinatubo eruption but then immediately followed by a rise, likely due to the volatile nature of subtropics. The few data points in the temperature drop make the changepoint detection very challenging as they simply behave like noise. In conclusion, we cannot assert that these three latitudes were not affected by the Mt. Pinatubo eruption, but the fingerprint of the eruption on those latitudes (if any) is brief and weak that can be easily masked by the natural fluctuation, based on the MERRA-2 data we used in our analyses.

\begin{figure}
\centering
\includegraphics[width=14.2cm]{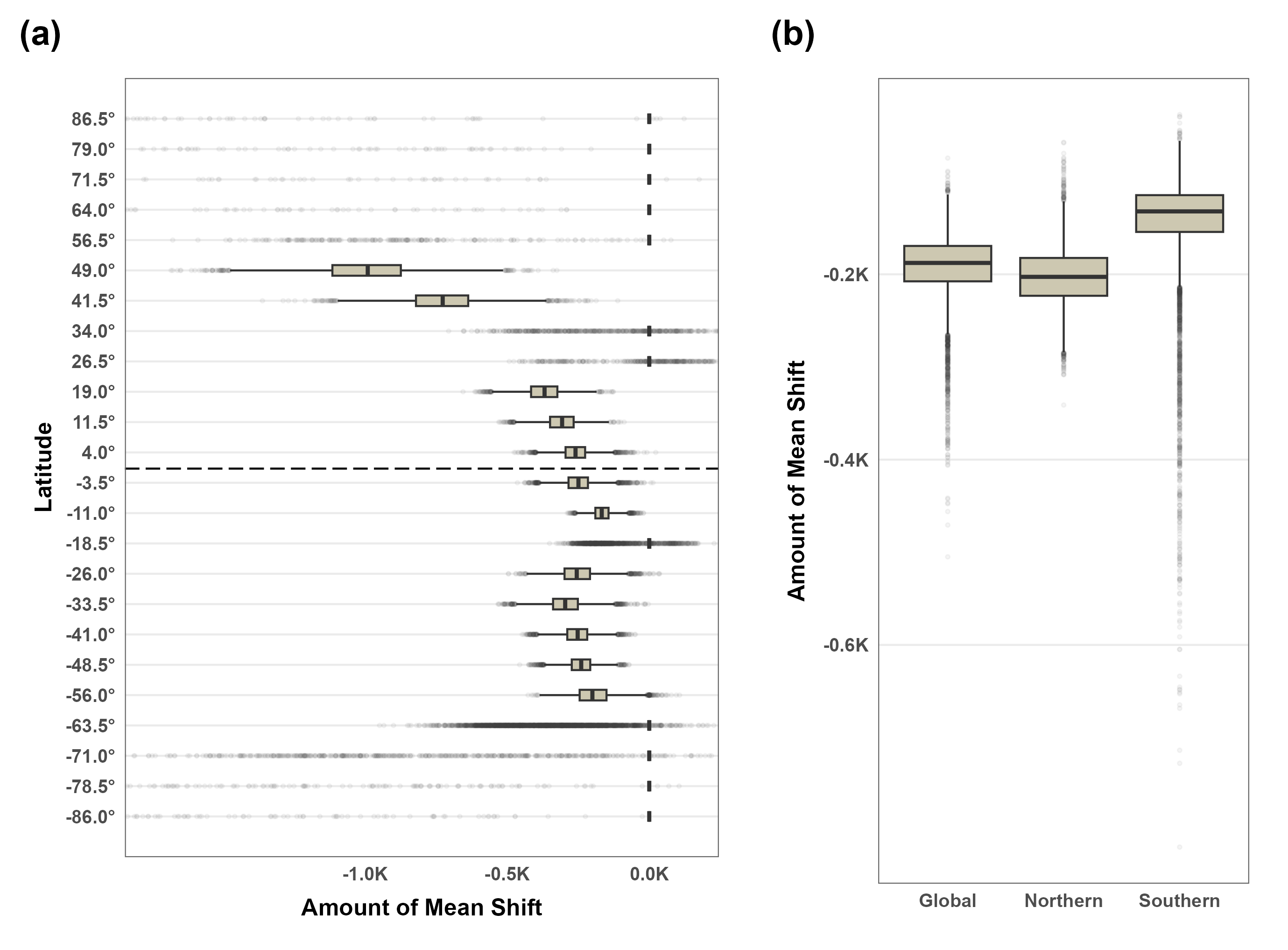}
\caption[Image]{Posterior distribution of $\gamma_0(\bs)$ (a) by latitude and (b) averaged over hemisphere after backtransforming to the original scale of surface temperature data. The dashed line in (a) marks the equator. }
\label{fig:mean_shift}
\end{figure}

The magnitude of the mean shift shown in Figure \ref{fig:surface_lat_mean} is not true to scale since we normalized the data to make the variance constant for all locations in the preprocessing step.  Figure \ref{fig:mean_shift} shows the posterior distribution of the actual mean shift at the detected latitudes, after back-transforming to match the scale of the original data. While there were more changepoints detected in the southern hemisphere, the magnitude of change is smaller compared to the northern hemisphere, and the change in global temperature is mostly driven by the latitudes $41.5^\circ N$ and $49^\circ N$. The average mean change in the southern hemisphere, northern hemisphere, and the globe is $-0.136^\circ C, -0.205^\circ C,$ and $-0.191^\circ C,$ respectively. Our findings are consistent with the existing studies \citep{self1993}, which report that the Mt. Pinatubo eruption led to a global cooling of $0.4^\circ C$ between years $1991$ and $1993$, driven mostly by a cooling of $0.5^\circ C$ in Northern Hemisphere. The estimated change amounts by our model are smaller in magnitude, which is expected since our method only measures the amount of mean shift at the time of the changepoint as opposed to the average change over two years. The negative $\gamma_{1F}$ estimate also indicates that the temperature continues to cool down after the changepoint.

Our findings from the analysis of stratospheric AOD and surface temperature data are consistent with the previous studies in general. However, our work complements the previous studies by providing quantitative regional measures of the impact of the Pinatubo eruption on aerosol and climate, as well as specifying the spatial pattern of the impact.

\section{Discussion}\label{sec:disc}

Motivated by the need to quantify the impact of the Mt. Pinatubo eruption on regional climate and specify its spatial pattern, we proposed a Bayesian framework to simultaneously detect and estimate changepoint for spatio-temporal data. Our approach can identify which spatial locations have a changepoint and, meantime, allows the changepoints and the magnitude of change to vary spatially, as opposed to assuming that all locations experience the same amount of change at the same time point. Furthermore, our model respects the diffusion pattern of the impact of the volcanic eruption and takes advantage of spatial correlation in changepoint detection and estimation. In particular, our method ensures that changepoints at locations other than the event origin occur strictly after the initial changepoint. This feature helps to prevent detecting changepoints caused by unrelated events, providing a more focused analysis of the targeted atmospheric event. 
The validity and effectiveness of our approach are demonstrated through simulations. By applying our method to AOD and surface temperature data, we successfully captured the spatial patterns of changepoints, which reveals the progression of injected aerosol and its spatially heterogeneous impact on regional aerosol optical depth and surface temperatures.

For convenience, we have opted for the exponential covariance function, a stationary model, to address spatial correlation in the data. However, given our large spatial domain, it may be more realistic to consider a nonstationary covariance function, such as those proposed by \cite{shand2017modeling}. 
Other possible improvements of covariance models include employing a chordal of circular Mat\'ern covariance function instead of an exponential covariance model \citep{guinness2016isotropic}, and a more flexible nonseparable or even asymmetric space-time covariance model \citep{gneiting2002nonseparable} for $U(\bs,t)$. Another interesting extension of our work is to directly consider a bivariate time series of AOD and surface temperature in changepoint detection and estimation. This may especially benefit the surface temperature due to the weaker signal in temperature than in AOD. This extension is expected to be involving, though, as the dependency structure of the bivariate data will be complex, and the changepoint processes for the two variables will also interact with each other. Additionally, our current method focuses on detecting the time at which the impact of the eruption first reaches each location. Capturing the whole life process of Pinatubo impact including when the data ``returns to normal”  will be an exciting but nontrial extention of our current method as that requires estimation of more than one changepoint at each location, aligning with the concept of ``epidemic changepoints'' \citep{tucker2023elastic}.
    
Finally, we by no means imply a causal relationship between the Mt. Pinatubo eruption and the detected changepoints. These changepoints could be due to other climate and weather events, such as the El Ni\~no effect, or other internal variations. Our aim in this article is to capture a plausible impact of the Mt. Pinatubo volcanic eruption. Separating Mt. Pinatubo's impact from other scenarios and establishing a pure causal relationship between the volcanic eruption and climate impact is nontrivial and is a topic we reserve for future studies.

\section*{Acknowledgments}

\paragraph{Funding Statement}
This paper describes objective technical results and analysis. Any subjective views or opinions that might be expressed in the paper do not necessarily represent the views of the U.S. Department of Energy or the United States Government.
Sandia National Laboratories is a multimission laboratory managed and operated by National Technology \& Engineering Solutions of Sandia, LLC, a wholly owned subsidiary of Honeywell International Inc., for the U.S. Department of Energy’s National Nuclear Security Administration under contract DE-NA0003525. SAND2024-XXXX. Li’s research is also partially supported by NSF-2124576. 

\bibliographystyle{unsrtnat}
\bibliography{sample.bib}       

\newpage
\appendix

\section{Supplement to ``Tracing the impacts of Mount Pinatubo eruption on regional climate using spatially-varying changepoint detection''}

\subsection{MCMC sampling}
We write the $\boldsymbol{\mu}_\tau$ as 
\begin{align*} 
\boldsymbol{\mu}_\tau &= \boldsymbol{\mu}_{F} + \boldsymbol{\mu}_{R} \circ \mathbf{1_\tau^+} \\
&= \mathbf{Z}_\tau \boldsymbol{\alpha} + \boldsymbol{\mu}_{R} \circ \mathbf{1_\tau^+}
\end{align*}
where $\boldsymbol{\mu}_R = \mathbf{\gamma_{0R}} + \mathbf{\gamma_{1R}}\cdot (\mathbf{t}\otimes \mathbf{1_N} - \mathbf{1_M}\otimes \boldsymbol{\tau})$, $\mathbf{Z}_\tau  = \begin{bmatrix}
    \mathbf{1_{MN}} & \mathbf{1_\tau^+} & (t\otimes\mathbf{1}_N - \mathbf{1}_M \otimes \boldsymbol\tau) \circ \mathbf{1_\tau^+}
\end{bmatrix},$ and each component of $\boldsymbol{\alpha} = \begin{bmatrix} \alpha_0 & \gamma_0 & \gamma_1 \end{bmatrix}^T$ represents the pre-changepoint mean, amount of mean shift, and post-changepoint temporal trend, respectively.

\begin{enumerate}[start=1,label={\bfseries Step \arabic* :},leftmargin=1.5cm]
    \item Sample $\boldsymbol{\mu}_\tau$ without conditioning on $\mathbf{U}.$
    \begin{enumerate}[(i)]
        \item Given prior $\boldsymbol{\alpha} \sim N(0,s^2_a I_3)$, the full conditional for $\boldsymbol{\alpha}$ with $\mathbf{U}$ integrated out is
\begin{align*}
    \boldsymbol{\alpha} & \mid \mathbf{Y},\boldsymbol{\tau}, \boldsymbol{\mu}_R, \theta  \sim N(\boldsymbol{\mu_\alpha},\boldsymbol{\Sigma_\alpha}),
\intertext{where} \boldsymbol{\Sigma_\alpha} & = \left({\mathbf{Z}^T(\sigma^2_1 \mathbf{I}_{MN} + \boldsymbol{\Sigma_U})^{-1}\mathbf{Z}}+ \frac{\mathbf{I}_3}{s^2_a} \right)^{-1} ,\\
 \boldsymbol{\mu_\alpha} &= \boldsymbol{\Sigma_\alpha} \mathbf{Z}^T (\sigma^2_1 \mathbf{I}_{MN} + \boldsymbol{\Sigma_U})^{-1} (\mathbf{Y}-(\boldsymbol{\epsilon}_\gamma + \boldsymbol{\mu}_{R})\circ \mathbf{1_\tau^+}).
\end{align*}
    \item Given prior $\tau_0 \sim Cat(\boldsymbol{\pi} = (\pi_1,\pi_2,\ldots , \pi_M)),$ the full conditional for $\tau_0$ with $\mathbf{U}$ integrated out is
    \begin{align*}
        \tau_0 &\mid \mathbf{Y}, \boldsymbol{\mu}_\tau, \boldsymbol{\tau},\theta \sim Cat(\boldsymbol{\widetilde{\pi}} = (\widetilde{\pi}_1, \ldots, \widetilde{\pi}_M)),
    \intertext{where} 
    \log \widetilde{\pi}_k \propto -\frac{1}{2}&(\mathbf{Z}_\tau \boldsymbol{\alpha} + (\boldsymbol{\epsilon}_\gamma + \boldsymbol{\mu}_{R})\circ \mathbf{1_\tau^+})^T(\sigma^2_1 \mathbf{I}_{MN} + \boldsymbol{\Sigma_U})^{-1}
    (\mathbf{Z}_\tau \boldsymbol{\alpha} + (\boldsymbol{\epsilon}_\gamma + \boldsymbol{\mu}_{R})\circ \mathbf{1_\tau^+}).
    \end{align*}
    \item See section \ref{mcmc} on how to sample $\boldsymbol{\Delta}$ and $\boldsymbol{\mu}_R$
     \end{enumerate}
    \item Sample $\mathbf{U}.$
    \begin{enumerate}[(i)]
        \item The full conditional for $\mathbf{U}$ is 
        \begin{align*}
            \mathbf{U} & \mid \cdot \sim N(\boldsymbol{\widetilde{\mu}_U},\boldsymbol{\widetilde{\Sigma}_U}),
            \intertext{where}
            \boldsymbol{\widetilde{\Sigma}_U} &= \frac{\mathbf{I}_{MN}}{\sigma^2_1} + \boldsymbol{\Sigma_U}^{-1},\\
            \boldsymbol{\widetilde{\mu}_U} &= \boldsymbol{\widetilde{\Sigma}_U}^{-1} \left(\frac{\mathbf{Y} - \mathbf{Z}_\tau \boldsymbol{\alpha} - (\boldsymbol{\epsilon}_\gamma + \boldsymbol{\mu}_{R})\circ \mathbf{1_\tau^+})}{\sigma^2_1} \right).
        \end{align*}
    \end{enumerate}
     \item Sample $\theta.$
    \begin{enumerate}[(i)]
        \item Given priors $\sigma^2_1 \sim IG(a_1,b_1), \sigma^2_\gamma \sim IG(a_2,b_2), \sigma^2_U \sim IG(a_3,b_3), \sigma^2_\Delta \sim IG(a_4,b_4),$ the full conditionals are
\begin{align*}
\sigma^2_1 \mid \cdot &\sim IG\left(a_1 + \frac{MN}{2}, b_1 + \frac{(\mathbf{Y}-\mathbf{U}-\boldsymbol{\mu}_\tau- \boldsymbol{\epsilon}_\gamma \circ \mathbf{1_\tau^+})^T(\mathbf{Y}-\mathbf{U}-\boldsymbol{\mu}_\tau - \boldsymbol{\epsilon}_\gamma \circ \mathbf{1_\tau^+})}{2} \right)\\
\sigma^2_\gamma \mid \cdot &\sim IG\left(a_2 + \frac{K}{2}, b_2 + \frac{\boldsymbol{\epsilon}_\gamma^T\boldsymbol{\epsilon}_\gamma}{2}\right)\\
\sigma^2_U \mid \cdot &\sim IG\left(a_3 + \frac{MN}{2}, b_3 + \frac{\mathbf{U}^T \mathbf{R}(\phi_U)^{-1}\otimes \mathbf{R}(\psi_U)^{-1} \mathbf{U}}{2} \right)\\
\sigma^2_\Delta \mid \cdot &\sim IG\left(a_4 + \frac{N}{2}, b_4 + \frac{1}{2}(\log \boldsymbol{\Delta} - \mathbf{X}\boldsymbol{\beta})^T \mathbf{R}(\psi_\Delta)^{-1} (\log\boldsymbol{\Delta} - \mathbf{X}\boldsymbol{\beta})\right)\\
\end{align*}
where $K = \text{number of post-cp indices}.$

 \item Given $\sigma^2_{\gamma_i} \mid \gamma_{iF} \sim TruncIG(\min=0, \max= \gamma_{iF}^2/9,a_5,b_5), \quad i \in \{0,1\},$ the full conditionals are
 \begin{align*}
     \sigma^2_{\gamma_0} \mid \cdot &\sim TruncIG\left (\min=0,\max=\frac{\gamma_{0F}^2}{9}, a_5 + \frac{N}{2}, b_5 + \frac{\boldsymbol{\gamma_{0R}}^T \mathbf{R}(\psi_{\gamma_{0R}})^{-1}\boldsymbol{\gamma_{0R}}}{2} \right)\\
     \sigma^2_{\gamma_1} \mid \cdot &\sim TruncIG\left (\min=0,\max=\frac{\gamma_{1F}^2}{9}, a_5 + \frac{N}{2}, b_5 + \frac{\boldsymbol{\gamma_{1R}}^T \mathbf{R}(\psi_{\gamma_{1R}})^{-1}\boldsymbol{\gamma_{1R}}}{2} \right)
 \end{align*}
\item Given prior $\boldsymbol{\beta} \sim N(\mathbf{0},s^2_b \mathbf{I}_2),$ the full conditional for $\boldsymbol{\beta}$ is
\begin{align*}
    \boldsymbol{\beta} &\sim N(\boldsymbol{\mu_\beta}, \boldsymbol{\Sigma_\beta}),
    \intertext{where}
    \boldsymbol{\Sigma_\beta} &= \left(\frac{\mathbf{X}^T \mathbf{R}(\psi_\Delta)^{-1} \mathbf{X}}{\sigma^2_\Delta} + \frac{\mathbf{I}_2}{s^2_b}\right)^{-1},\\
    \boldsymbol{\mu_\beta} &=\boldsymbol{\Sigma_\beta} \left(\frac{\mathbf{X}^T \mathbf{R}(\psi_\Delta)^{-1}\log \boldsymbol{\Delta}}{\sigma^2_\Delta}\right).
\end{align*}

\item The full conditional for $\boldsymbol{\epsilon}_\gamma$ is
\begin{align*}
    \boldsymbol{\epsilon}_\gamma &\sim N(\boldsymbol{\widetilde{\mu}_\gamma},\boldsymbol{\widetilde{\Sigma} _\gamma}),
    \intertext{where} \boldsymbol{\widetilde{\Sigma} _\gamma} &= diag\left( \left(\frac{\mathbf{1_\tau^+}}{\sigma^2_1} + \frac{\mathbf{1}}{\sigma^2_\gamma}\right)^{-1} \right),\\
    \boldsymbol{\widetilde{\mu}_\gamma}&= \boldsymbol{\widetilde{\Sigma} _\gamma}\left( \frac{Y-\mu_\tau - U}{\sigma^2_1}\right) \circ \mathbf{1_\tau^+}.
\end{align*}
\item Given priors $\phi_U \sim Unif(l_1,u_1), \psi_U\sim Unif(l_2,u_2), \psi_\Delta \sim Unif(l_3,u_3),$ the full conditionals are
\begin{align*}
    f(\phi_U \mid \cdot) & \propto |\mathbf{R}(\phi_U)|^{-N/2}\exp\left(-\frac{\mathbf{U}^T(\mathbf{R}(\phi_U)^{-1} \otimes \mathbf{R}(\psi_U)^{-1})\mathbf{U}}{2\sigma^2_U}\right) \mathds{1}(l_1 < \phi_U < u_1)\\
    f(\psi_U \mid \cdot) & \propto |\mathbf{R}(\psi_U)|^{-M/2}\exp\left(-\frac{\mathbf{U}^T(\mathbf{R}(\phi_U)^{-1} \otimes \mathbf{R}(\psi_U)^{-1})\mathbf{U}}{2\sigma^2_U}\right) \mathds{1}(l_2 < \psi_U < u_2) \\
    f(\psi_\Delta \mid \cdot) & \propto |\mathbf{R}(\psi_\Delta)|^{-1/2}\exp\left(-\frac{(\log\boldsymbol{\Delta} - \mathbf{X}\boldsymbol{\beta})^T\mathbf{R}(\psi_\Delta)^{-1}(\log\boldsymbol{\Delta} - \mathbf{X}\boldsymbol{\beta})}{2\sigma^2_\Delta}\right)\\
    & \cdot \mathds{1}(l_3 < \psi_\Delta < u_3)\\
\end{align*} 
These full conditional distributions do not have a closed form. We use a Metropolis Hastings algorithm to sample these parameters. The proposal densities are 
\begin{align*}
    q(\phi_U^*\mid\phi_U^{k-1}) &\sim N(\phi_U^{k-1},s_1^2)\mathds{1}(l_1<\phi_U^*<u_1)\\
    q(\psi_U^*\mid\psi_U^{k-1}) &\sim N(\psi_U^{k-1},s_s^2)\mathds{1}(l_2<\psi_U^*<u_2)\\
    q(\psi_\Delta^*\mid\psi_\Delta^{k-1}) &\sim N(\psi_\Delta^{k-1},s_3^2)\mathds{1}(l_3<\psi_\Delta^*<u_3)\\
\end{align*}
The acceptance ratios are 
\begin{align*}
    r_{\phi_U} &= \left(\frac{|\mathbf{R}(\phi_U^*)|}{|\mathbf{R}(\phi_U^{k-1}|}\right)^{-N/2}\exp\left(-\frac{\mathbf{U}^T( (\mathbf{R}(\phi_U^*)^{-1}-\mathbf{R}(\phi_U^{k-1})^{-1}) \otimes \mathbf{R}(\psi_U)^{-1} )\mathbf{U}}{2\sigma^2_U}\right)\\ 
    &\phantom{=} \cdot \left(\frac{\Phi(\frac{u_1 - \phi_U^{k-1}}{s_1}) - \Phi(\frac{l_1-\phi_U^{k-1}}{s_1})}{\Phi(\frac{u_1 - \phi_U^*}{s_1}) - \Phi(\frac{l_1 - \phi_U^*}{s_1})}\right) \\
    r_{\psi_U} &= \left(\frac{|\mathbf{R}(\psi_U^*)|}{|\mathbf{R}(\psi_U^{k-1}|}\right)^{-M/2}\exp\left(-\frac{\mathbf{U}^T( \mathbf{R}(\phi_U)^{-1} \otimes (\mathbf{R}(\psi_U^*)^{-1}-\mathbf{R}(\psi_U^{k-1})^{-1}) )\mathbf{U}}{2\sigma^2_U}\right)\\ 
    &\phantom{=} \cdot \left(\frac{\Phi(\frac{u_2 - \psi_U^{k-1}}{s_2}) - \Phi(\frac{l_2-\psi_U^{k-1}}{s_2})}{\Phi(\frac{u_2 - \psi_U^*}{s_2}) - \Phi(\frac{l_2 - \psi_U^*}{s_2})}\right) \\
    r_{\psi_\Delta} &= \left(\frac{|\mathbf{R}(\psi_\Delta^*)|}{|\mathbf{R}(\psi_\Delta^{k-1}|}\right)^{-\frac{1}{2}}\\
    &\phantom{=} \cdot \exp\left(-\frac{(\log\boldsymbol{\Delta} - \mathbf{X}\boldsymbol{\beta})^T (\mathbf{R}(\psi_\Delta^*)^{-1}-\mathbf{R}(\psi_\Delta^{k-1})^{-1}) (\log\boldsymbol{\Delta} - \mathbf{X}\boldsymbol{\beta})}{2\sigma^2_\Delta}\right)\\ 
     &\phantom{=} \cdot \left(\frac{\Phi(\frac{u_3 - \psi_\Delta^{k-1}}{s_3}) - \Phi(\frac{l_3-\psi_\Delta^{k-1}}{s_3})}{\Phi(\frac{u_3 - \psi_\Delta^*}{s_3}) - \Phi(\frac{l_3 - \psi_\Delta^*}{s_3})}\right) \\
\end{align*}
    \end{enumerate}
\end{enumerate}

\subsubsection{Adaptive Metropolis Hastings} \label{mcmc}
To sample $\boldsymbol{\Delta},$ we draw $\log \boldsymbol{\Delta}$ and take the exponential. The log of the full conditional distribution for $\log \boldsymbol{\Delta}$ with $\mathbf{U}$ integrated out is
\begin{align*}
\ell(\log \boldsymbol{\Delta} \mid \cdot) \propto  & -\frac{\log\boldsymbol{\Delta}^T \mathbf{R}(\psi_\Delta)^{-1}\log\boldsymbol{\Delta} - 2\log\boldsymbol{\Delta}^T \mathbf{R}(\psi_\Delta)^{-1}\mathbf{X}\boldsymbol{\beta}}{2\sigma^2_\Delta}\\
& - \frac{1}{2}(\mathbf{Z}_\tau \boldsymbol{\alpha} + (\boldsymbol{\epsilon}_\gamma + \boldsymbol{\mu}_{R})\circ \mathbf{1_\tau^+})^T(\sigma^2_1 \mathbf{I}_{MN} + \boldsymbol{\Sigma_U})^{-1}(\mathbf{Z}_\tau \boldsymbol{\alpha} + (\boldsymbol{\epsilon}_\gamma + \boldsymbol{\mu}_{R})\circ \mathbf{1_\tau^+}) .
\end{align*}

We sample $\boldsymbol{\Delta} = (\Delta_{s_1}, ..., \Delta_{s_N})^T$ using adaptive component-wise Metropolis Hastings algorithm as follows:
\begin{enumerate}[leftmargin=0.5cm]
    \item Set batch size $B.$ Initialize the log proposal variances $ls_i$ for each spatial components. 
    \item For $k = 1,..., K:$
\begin{enumerate}
        \item For $i = 1,..., N:$
        \begin{enumerate}
            \item Draw $\log \Delta^*_{s_i}$ from proposal density \[q(\log\Delta^*_{s_i} \mid \log\Delta^{k-1}_{s_i}) \sim N(\log \Delta_{s_i}^{k-1},\exp(ls_i)).\]
            \item Calculate acceptance ratio 
            \begin{align*}
              r = \exp(-&(\log\boldsymbol{\Delta}^*-\log\boldsymbol{\Delta}^{k-1})^T \mathbf{R}(\psi_\Delta)^{-1}(\log\boldsymbol{\Delta}^* + \log\boldsymbol{\Delta}^{k-1} - \mathbf{X}\boldsymbol{\beta}) /2\sigma^2_\Delta + \\ 
            &(\boldsymbol{\mu_{\tau}}^* + \boldsymbol{\mu_{\tau}}^{k-1} + \boldsymbol{\epsilon}_\gamma \circ (\mathbf{1_\tau^+}^* + \mathbf{1_\tau^+}^{k-1}) - 2(\mathbf{Y}-\mathbf{U}))^T\\
            &(\boldsymbol{\mu_{\tau}}^* - \boldsymbol{\mu_{\tau}}^{k-1} + \boldsymbol{\epsilon}_\gamma \circ (\mathbf{1_\tau^+}^* - \mathbf{1_\tau^+}^{k-1})/2\sigma^2_1  ) ), 
            \intertext{where}
            \boldsymbol{\Delta}^* &= (\Delta_{s_1}^k,...,\Delta_{s_{i-1}}^k,\Delta^*_{s_i},\Delta_{s_{i+1}}^{k-1},...,\Delta_{s_N}^{k-1})^T,
            \intertext{and}
            \mu_\tau^*(\mathbf{s},t) &= \alpha_0 + [\gamma_0(\mathbf{s}) + \gamma_1(\mathbf{s})\cdot (t-\tau_0-\Delta^*(\mathbf{s})) ]\cdot \mathds{1}(t > \tau_0 + \Delta^*(\mathbf{s})).
            \end{align*}
            \item Sample $u \sim Unif(0,1).$ Set $\Delta ^k = \begin{cases} \Delta^* & \text{if } r > u \\ \Delta^{k-1} & o.w. \end{cases}.$ Set $a^k = \begin{cases} 1 & \text{if } r > u \\ 0 & o.w. \end{cases}.$
        \end{enumerate}
        \phantom{ljk}
        \item If $k = jB$ for integer $j\geq 1:$ 
        \begin{enumerate}
            \item Calculate the average acceptance rate for the $j^\textsuperscript{th}$ batch as $a_j = \sum\limits_{l=k - B +1}^{k} a^{l}/B.$ 
            \item Update $ls_i = \begin{cases} ls_i + \Delta(j), & a_j > 0.44\\
        ls_i - \Delta(j), & a_j \leq 0.44,\end{cases}$ \\ where $\Delta(j) = \min(0.1,1/\sqrt{j}).$
        \end{enumerate}
    \end{enumerate}
\end{enumerate}

\bigskip
\bigskip

The full conditional for $\boldsymbol{\gamma_{0R}}$ is
\begin{align*}
    \mathbf{1}_M \otimes \boldsymbol{\gamma_{0R}} &\sim N( \boldsymbol{\widetilde{\mu}_{\gamma_0}}, \boldsymbol{\widetilde{\Sigma}_{\gamma_0}}),
    \intertext{where}
    \boldsymbol{\widetilde{\Sigma}_{\gamma_0}} &= \frac{diag(\mathbf{1_\tau^+})}{\sigma^2_1} + \mathbf{I}_N\otimes \boldsymbol{\Sigma_{\gamma_0}},\\
    \boldsymbol{\widetilde{\mu}_{\gamma_0}} &=  \boldsymbol{\widetilde{\Sigma}_{\gamma_0}}^{-1}\left(\frac{diag(\mathbf{1_\tau^+})(\mathbf{Y}-\boldsymbol{\mu}_\tau - \boldsymbol{\epsilon}_\gamma)}{\sigma^2_1}\right).
\end{align*}
Since $diag(\mathbf{1_\tau^+})$ is not separable, there is no easy way to sample from this distribution directly. Thus, we use the adaptive component-wise Metropolis Hastings to update $\boldsymbol{\gamma_{0R}}$ instead of using a Gibbs Sampler. The steps for updating $\boldsymbol{\gamma_{0R}}$ and $\boldsymbol{\gamma_{1R}}$ are same as $\boldsymbol{\Delta},$ where the acceptance ratio for $\boldsymbol{\gamma_{0R}}$ is given by 
\begin{align*}
    r = \exp(&-(\boldsymbol{\gamma_{0R}}^* + \boldsymbol{\gamma_{0R}}^{k-1})^T \boldsymbol{\Sigma_{\gamma_0}}^{-1} (\boldsymbol{\gamma_{0R}}^* - \boldsymbol{\gamma_{0R}}^{k-1})/2 + \\
    & ( (\mathbf{1}_M \otimes (\boldsymbol{\gamma_{0R}}^* + \boldsymbol{\gamma_{0R}}^{k-1})) \circ \mathbf{1_\tau^+} - 2\mathbf{A})^T( (\mathbf{1}_M \otimes (\boldsymbol{\gamma_{0R}}^* - \boldsymbol{\gamma_{0R}}^{k-1})) \circ \mathbf{1_\tau^+})/2\sigma^2_1 ),
    \intertext{where}\mathbf{A} &= \mathbf{Y} - \mathbf{U} - \boldsymbol{\mu}_F - \boldsymbol{\gamma_{1R}}\circ (\mathbf{t}\otimes \mathbf{1_N} - \mathbf{1_M}\otimes \boldsymbol{\tau}),
\end{align*}
and the acceptance ratio for $\boldsymbol{\gamma_{1R}}$ is given by 
\begin{align*}
    r = .\exp(- &(\boldsymbol{\gamma_{1R}}^* +\boldsymbol{\gamma_{1R}}^{k-1})^T \boldsymbol{\Sigma_{\gamma_1}}^{-1} (\boldsymbol{\gamma_{1R}}^* - \boldsymbol{\gamma_{1R}}^{k-1})/2 + \\
    & ( (\mathbf{1}_M \otimes (\boldsymbol{\gamma_{1R}}^* + \boldsymbol{\gamma_{1R}}^{k-1})) \circ (\mathbf{t}\otimes \mathbf{1_N} - \mathbf{1_M}\otimes \boldsymbol{\tau}) \circ \mathbf{1_\tau^+} - 2\mathbf{A})^T \\
    & ( (\mathbf{1}_M \otimes (\boldsymbol{\gamma_{1R}}^* - \boldsymbol{\gamma_{1R}}^{k-1})) \circ (\mathbf{t}\otimes \mathbf{1_N} - \mathbf{1_M}\otimes \boldsymbol{\tau}) \circ \mathbf{1_\tau^+})/2\sigma^2_1 ),
\intertext{where} \mathbf{A} &= \mathbf{Y} - \mathbf{U} - \boldsymbol{\mu}_F - \boldsymbol{\gamma_{0R}}.
\end{align*}

\section{Tables and Figures}

\begin{figure}
\centering
\includegraphics[width=14.2cm]{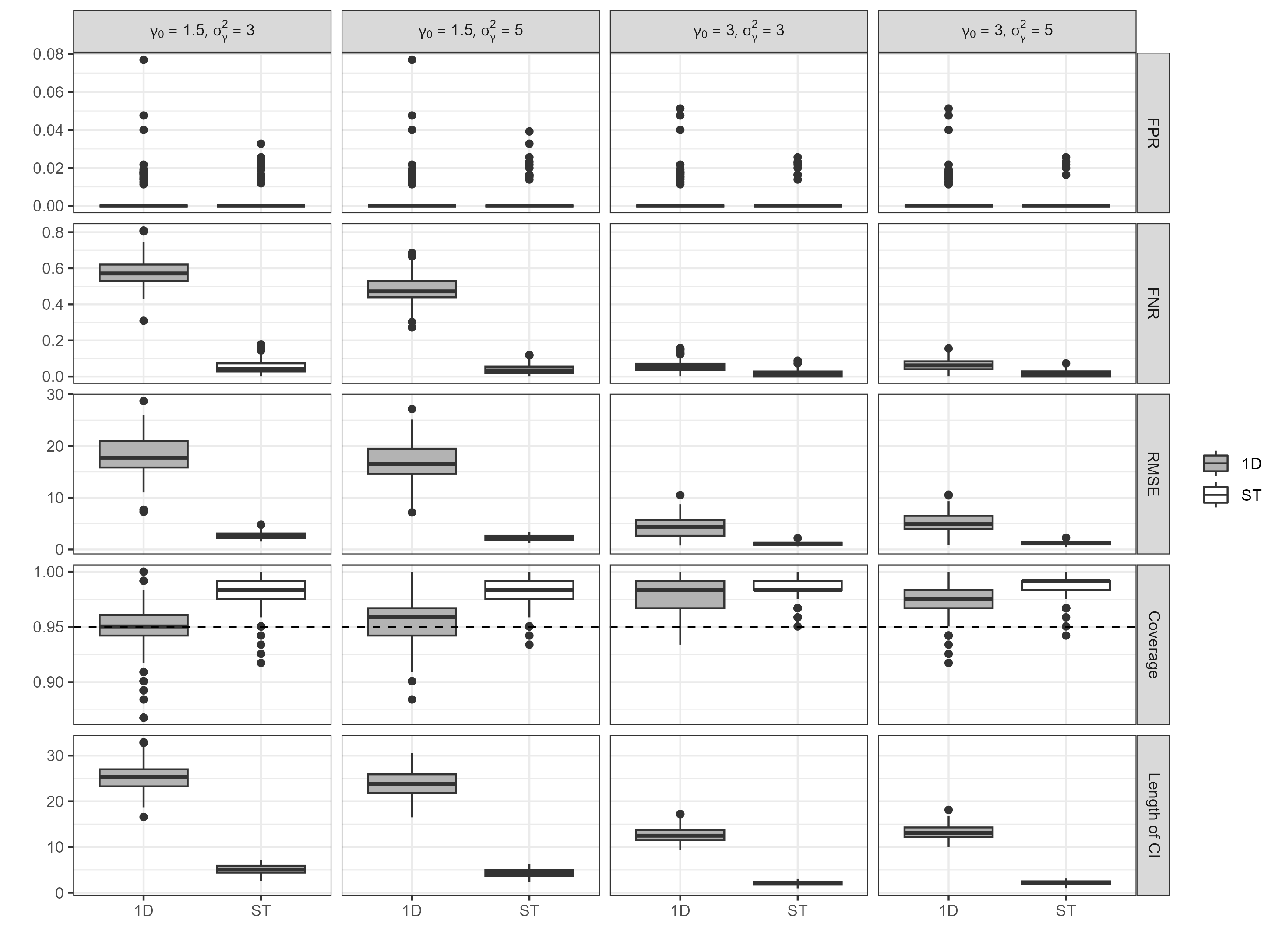}
\caption{Boxplots of RMSE, FPR, FNR, the empirical coverage probability  and the length of 95\% credible intervals under $\gamma_0 \in \{1.5,3\}$ and $\sigma^2_\gamma \in \{3,5\}$ . “ST” is our spatio-temporal model and “1D” is the univariate method. }
\label{fig:sim}
\end{figure}

\begin{table}[h!]
\centering
\begin{tabular}{clcc}
\hline
Latitude & Posterior Mode & $\mathbb{P}$(no changepoint) & 95\% CI \\\hline
$86.0^\circ$S & Dec 1995 (No CP) & 0.9905 & (Dec 1995, Dec 1995) \\
$78.5^\circ$S & Dec 1995 (No CP) & 0.9888 & (Dec 1995, Dec 1995) \\
$71.0^\circ$S & Dec 1995 (No CP) & 0.9762 & (Dec 1995, Dec 1995) \\
$63.5^\circ$S & Dec 1995 (No CP) & 0.7955 & (Apr 1992, Dec 1995) \\
$56.0^\circ$S & Mar 1992 & 0.0037 & (Jul 1991, Jul 1994) \\
$48.5^\circ$S & Oct 1991 & 0 & (Jul 1991, Feb 1992)\\
$41.0^\circ$S & Sep 1991 & 0 & (Jul 1991, May 1992)\\
$33.5^\circ$S & Sep 1991 & 0.0001 & (Jul 1991, Jan 1992) \\
$26.0^\circ$S & Oct 1991 & 0.0112 & (Jun 1991, Nov 1992)\\
$18.5^\circ$S & Dec 1995 (No CP) & 0.8470 & (Apr 1995, Dec 1995) \\
$11.0^\circ$S & Sep 1991 & 0 & (Aug 1991, Aug 1992)\\
$3.5^\circ$S & Apr 1992 & 0 & (Aug 1991, Sep 1991)\\
$4.0^\circ$N & May 1992 & 0 & (Oct 1991, Aug 1992) \\
$11.5^\circ$N & Sep 1991 & 0 &(Jul 1991, Dec 1991)\\
$19.0^\circ$N & Sep 1991 & 0& (Jun 1991, Nov 1991)\\
$26.5^\circ$N & Dec 1995 (No CP) & 0.9878 & (Dec 1995, Dec 1995)\\
$34.0^\circ$N & Dec 1995 (No CP) & 0.9763 & (Dec 1995, Dec 1995)\\
$41.5^\circ$N & Sep 1991 & 0 & (Jul 1991, Jun 1992)\\
$49.0^\circ$N & Sep 1991 & 0 & (Jul 1991, Mar 1992)\\
$56.5^\circ$N & Dec 1995 (No CP) & 0.9788 & (Dec 1995, Dec 1995)\\
$64.0^\circ$N & Dec 1995 (No CP) & 0.9986 & (Dec 1995, Dec 1995)\\
$71.5^\circ$N & Dec 1995 (No CP) & 0.9984 & (Dec 1995, Dec 1995)\\
$79.0^\circ$N & Dec 1995 (No CP) & 0.9943 & (Dec 1995, Dec 1995)\\
$86.5^\circ$N & Dec 1995 (No CP) & 0.9984 & (Dec 1995, Dec 1995)\\
\hline
\end{tabular}
\caption{Summary of detected changepoints for surface temperature}
\label{tab:surface_temp_res}
\end{table}

\begin{figure}
\centering
\includegraphics[width=14.2cm]{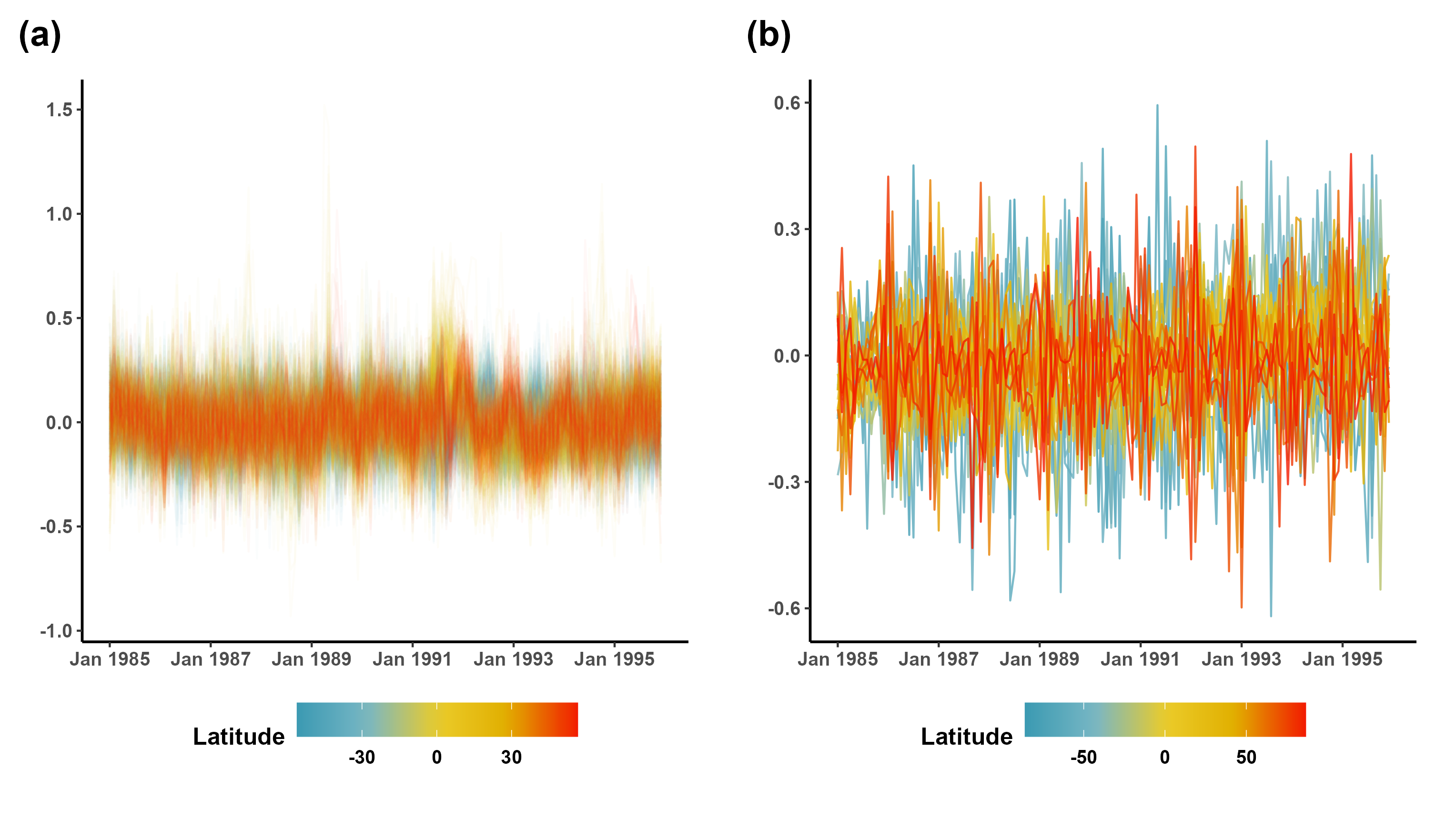}
\caption{Residual plot of (a) stratospheric AOD and (b) surface temperature under $\sigma^2_1 = \sigma^2_2$}
\label{fig:residual}
\end{figure}

\end{document}